\documentclass[ALICE,manyauthors]{cernphprep}

\newcommand{\GeVc}{\ensuremath{\textrm{GeV}/c}\xspace}
\newcommand{\pt}{\ensuremath{p_{\rm T}}\xspace}

\newcommand{\dEdx}{\ensuremath{\textrm{d}E/\textrm{d}x}\xspace}

\newcommand{\RpPb}{\ensuremath{R_{\rm pPb}}\xspace}

\newcommand{\pPb}{p--Pb~}

\newcommand{\sqpA}{\ensuremath{\sqrt{s_{\rm NN}}} = 5.02~TeV\xspace}
\newcommand{\smb}{\ensuremath{\sigma_{\rm MB}^{\rm V0}}\xspace}
\newcommand{\smbeq}{\sigma_{\rm MB}^{\rm V0}}
\newcommand{\QpPb}{\ensuremath{Q_{\rm pPb}}\xspace}
\newcommand{\QCP}{\ensuremath{Q_{\rm cp}}\xspace}

\usepackage{graphicx}
\pdfoutput=1
\usepackage{color}
\usepackage[comma,square,numbers,sort&compress]{natbib}
\usepackage{siunitx}
\usepackage{hyperref}
\usepackage{placeins}
\usepackage{multirow}

\usepackage[utf8]{inputenc} 

\begin{document}

\begin{titlepage}
\PHyear{2019}
\PHnumber{250}      
\PHdate{28 October}  
%

\title{ Measurement of electrons from heavy-flavour hadron decays as a function of multiplicity
 in \pPb collisions at $\sqrt{s_{\rm NN}}$ = 5.02 TeV }
\ShortTitle{Electrons from heavy-flavour hadron decays as a function of multiplicity}   

\Collaboration{ALICE Collaboration\thanks{See Appendix~\ref{app:collab} for the list of collaboration members}}
\ShortAuthor{ALICE Collaboration} 

\begin{abstract}
The multiplicity dependence of electron production from heavy-flavour hadron decays as a function
 of transverse momentum was measured in {\mbox{p--Pb}} collisions at \sqpA  using the ALICE
 detector at the LHC. The measurement was performed in the centre-of-mass rapidity interval $-1.07 < y_{\rm cms} < 0.14$
 and transverse momentum interval 2 $<\pt<$ 16 ~\GeVc. The multiplicity dependence
 of the production of electrons from heavy-flavour hadron decays was studied by comparing the \pt spectra
 measured for different multiplicity classes with those measured in pp collisions (\QpPb) and in peripheral {\mbox{p--Pb}}
 collisions (\QCP). The \QpPb results obtained are consistent with unity within uncertainties in the measured \pt interval
 and event classes. This indicates that heavy-flavour decay electron production is consistent
 with binary scaling and independent of the geometry of the collision system. 
 Additionally, the results suggest that cold nuclear matter effects are negligible within uncertainties, in the production of heavy-flavour decay electrons at midrapidity in {\mbox{p--Pb}} collisions.

\end{abstract}
\end{titlepage}
\setcounter{page}{2}

\section{Introduction}\label{introduction}
Ultra-relativistic heavy-ion collisions provide suitable conditions to investigate the properties of strongly-interacting matter under extreme temperature and/or energy density. Under these conditions, lattice quantum chromodynamics calculations predict a transition from a hadronic to a partonic phase, known as the Quark-Gluon Plasma (QGP)~\cite{Bjorken:1982qr, Karsch:2001cy}.

Heavy quarks, i.e., charm and beauty quarks, are sensitive probes of the QGP as they are predominantly produced in the early stages of the collisions via hard scattering processes characterised by time scales shorter than the production time of the QGP~\cite{Andronic:2015wma, Liu:2012ax}. Since the heavy quark production and annihilation rates in the thermal phase are negligible~\cite{BraunMunzinger:2007tn}, they experience the entire space--time evolution of the system by interacting via elastic and radiative processes ~\cite{Baier:1996kr, Wicks:2007am,Braaten:1991jj}.

The nuclear modification factor ($R_{\rm AA}$) is commonly used to study the energy loss of partons in the medium. 
The $R_{\rm AA}$ is defined as the ratio between the transverse momentum (\pt) differential yield of the produced particles in nucleus-nucleus collisions and the \pt-differential cross section in proton-proton collisions, scaled by the average number of binary nucleon--nucleon collisions calculated with the Glauber model ~\cite{Miller:2007ri, Loizides:2017ack}.
In central Au--Au collisions at $\sqrt{s_{\rm NN}}$~= 200 GeV, 
both the production of charm mesons and electrons from heavy-flavour hadron decays are found to be  suppressed by a factor of 5 ($R_{\rm AA}\sim$  0.2) at midrapidity for $\pt>$ 3 \GeVc and $\pt>$ 5 \GeVc, respectively ~\cite{Adamczyk:2014uip, Abelev:2006db, Adare:2010de}.

In Pb--Pb collisions at $\sqrt{s_{\textrm{NN}}}=$ 2.76 and 5.02~TeV,  a similar suppression was observed not only 
for particles containing charm quarks, but also for those coming from beauty quark fragmentation
(B mesons and non-prompt J/$\psi$) ~\cite{Abelev:2012qh, Chatrchyan:2012np, Adam:2015rba, Adam:2015nna, Adam:2016khe, Aaboud:2018bdg, Acharya:2018hre}. Also, it was found that the production of jets from beauty quark fragmentation was strongly suppressed~\cite{Chatrchyan:2013exa}. The $R_{\rm AA}$ is about 0.4 for the jets associated to beauty quarks of the \pt range of 80--250 \GeVc for central Pb--Pb collisions at $\sqrt{s_{\textrm{NN}}}=$ 2.76~TeV.

The production of heavy quarks in heavy-ion collisions can be modified by initial-state effects in Cold Nuclear Matter (CNM), as well as by final-state effects i.e., energy loss in the dense medium. The CNM effects include the modification of the Parton Distribution Functions (PDFs) of the nuclei with respect to a superposition of nucleon PDFs, addressed by nuclear shadowing models~\cite{Eskola:2009uj, Helenius:2012wd} or gluon saturation models such as the Colour Glass Condensate (CGC) effective theory~\cite{Fujii:2013yja, Albacete:2012xq}. Furthermore, CNM effects also include Cronin-like enhancement ($k_{\rm T}$ broadening)~\cite{Lev:1983hh,  Wang:1998ww, Kopeliovich:2002yh} and energy loss in the initial~\cite{Vitev:2007ve} and final~\cite{Arleo:2010rb} stages of the collision. 

Initially, it was assumed that a QGP is not formed in proton--nucleus (p--A) collisions, so these collisions were used as a baseline for measurements in A--A collisions to test for possible CNM effects. The ALICE Collaboration reported the \pt-differential nuclear modification factor \RpPb of D mesons~\cite{Abelev:2014hha, Acharya:2019mno} and electrons from heavy-flavour hadron decay~\cite{Adam:2015qda} measured at midrapidity in \pPb collisions at $\sqrt{s_{\rm NN}}$~= 5.02~TeV.
The \RpPb at midrapidity are consistent with unity and with theoretical calculations including CNM effects, indicating that CNM effects are small in this kinematic region.
The \RpPb measured for B mesons~\cite{Khachatryan:2015uja} and jets from from beauty quark fragmentation~\cite{Khachatryan:2015sva} are also consistent with unity. 
All of these results indicate that initial-state effects are small for heavy-flavour production at midrapidity and, on their own, cannot explain the strong suppression observed at high \pt in nucleus-nucleus collisions.
However, at forward and backward rapidity, this scenario can be different:
muons from heavy-flavour hadron decays were measured by ALICE in \pPb collisions at $\sqrt{s_{\rm{NN}}}$~= 5.02~TeV~\cite{Acharya:2017hdv}  and by the PHENIX experiment
 in d$-$Au collisions at $\sqrt{s_{\rm{NN}}}$~= 200 GeV~\cite{Adare:2013lkk}. Both results show a small enhancement at backward rapidities which implies that CNM effects are present. At forward rapidities, the PHENIX results show a suppression, while at LHC energies, the ALICE results are compatible with unity. Similar results are also observed for prompt $\rm D^{0}$ measurements by the LHCb experiment for 0 $< \pt <$ 8 \GeVc~\cite{LHCb:2016huj}. 
The enhancement observed at backward rapidity is described by incoherent multiple scattering effects of partons in the Pb nucleus in the initial- and final-state interactions~\cite{Kang:2014hha}. The suppression observed by PHENIX at forward rapidity can be explained by gluon shadowing and/or energy loss in CNM~\cite{Vitev:2007ve}. Thus, at RHIC energies, the CNM effects at forward rapidity are important to describe the suppression observed in Au$+$Au collisions.

On the other hand, recent observations indicate that there may be collective effects in p--A collisions along with modifications observed in heavy-flavour production.
The nuclear modification factor of electrons from heavy-flavour hadron decays at midrapidity was found to be larger than unity in central d-Au collisions at $\sqrt{s_{\rm NN}}$~= 200 GeV in the transverse momentum interval 1.5 $< \pt <$ 5 \GeVc, measured by PHENIX~\cite{Adare:2012yxa} and the results are consistent with a model that includes radial flow effects~\cite{Sickles:2013yna}.
A positive value of the anisotropic flow parameter, $v_{2}$, for electrons~\cite{Acharya:2018dxy} and muons~\cite{Adam:2015bka} from heavy-flavour hadron decays was also observed in p--Pb collisions at $\sqrt{s_{\rm NN}}$~= 5.02~TeV. The results suggest that a collective behaviour induced via final-state effects may be present in small systems.

Measurements of the heavy-flavour particle multiplicity as a function of the number of charged-particle production in \pPb collisions can give more insight into the CNM effects, and possible final-state effects in small systems. 
These measurements might help to constrain the dependence of heavy-flavour production on the collision geometry and on the density of final-state particles, because Cronin-like enhancement due to multiple-parton scattering was observed to be stronger in central collisions than in peripheral collisions~\cite{Adam:2014qja}.

Final-state effects, energy loss, and collective behaviour are also sensitive to the particle multiplicity.
In Pb--Pb collisions, the suppression of $\rm D$ mesons and electrons from heavy-flavour hadron decays is stronger in central collisions than in peripheral collisions~\cite{Adam:2015nna, Adam:2016khe}.
The enhancement of electrons from heavy-flavour hadron decays in d--Au collisions is reproduced by a model that includes radial flow effects~\cite{Sickles:2013yna} and it is more pronounced in central collisions~\cite{Adare:2012yxa}.
Thus, if final-state effects are also present in p--Pb collisions, modification of the momentum distribution of heavy-flavour production could be expected in high-multiplicity p--Pb collisions.
Recently, ALICE measured the \pt-differential nuclear modification factor of D mesons for different multiplicity classes at midrapidity in \pPb collisions at $\sqrt{s_{\rm NN}}$~= 5.02~TeV~\cite{Adam:2016mkz, Acharya:2019mno}. These works have shown that the D-meson results are consistent with binary collision scaling of the yield in pp collisions, within the statistical and systematic uncertainties.

In this paper, the \pt-differential invariant cross section of electrons from heavy-flavour hadron decays produced in \pPb collisions at $\sqrt{s_{\rm NN}}$~= 5.02~TeV is measured both for minimum-bias collisions and for different charged-particle multiplicity classes.
This analysis extends the previously measured electron spectrum~\cite{Adam:2015qda} up to a \pt of 20 \GeVc which allows for the study of beauty production in p--Pb collisions, as beauty decays are the dominant source of electron production at $p_{\textrm{T}} >$ 4 \GeVc ~\cite{Abelev:2012sca}.

The nuclear modification factor of electrons from heavy-flavour hadron decays was calculated as

\begin{equation}
\RpPb= \frac{1}{\langle T_{\rm pPb}\rangle}\frac{ {\rm d}N^{ \rm  pPb}/{\rm d}p_{\rm T}}{{\rm d} \sigma^{\rm pp}/{\rm d}p_{\rm T}},
\end{equation} 

where $ \langle T_{\rm pPb}\rangle$  is the average nuclear overlap function,
$ {\rm d}N^{\rm pPb}/{\rm d}p_{\rm T}$ is the yield of electrons from heavy-flavour hadron decays in p--Pb collisions at $\sqrt{s_{\rm NN}}$~= 5.02~TeV, and ${\rm d}\sigma^{\rm pp}/{\rm d}p_{\rm T}$ is the cross section of electrons from heavy-flavour hadron decays in pp collisions  at $\sqrt{s}$~= 5.02~TeV.
The calculation of $ \langle T_{\rm pPb}\rangle$ using a Glauber model is discussed in Sec.~\ref{Event_selection}.

The multiplicity dependence of the electrons from heavy-flavour hadron decays was evaluated by means of the \QpPb factor, which is obtained by calculating the ratio of spectra in different multiplicity classes with respect to spectra in pp collisions, scaled by the number of binary nucleon--nucleon collisions: 

\begin{equation}
Q_{\rm  pPb} = \frac{1}{\langle T_{\rm pPb}^{\rm mult} \rangle}\frac{ {\rm d}N^{ \rm  pPb}_{\rm mult}/{\rm d}p_{\rm T}}{{\rm d} \sigma^{\rm pp}/{\rm d}p_{\rm T}},
\label{eq:QpPb}
\end{equation}

where $ \langle T_{\rm pPb}^{\rm mult} \rangle$ is the average nuclear overlap function in a given multiplicity class. The ${\rm d}N^{\rm pPb}_{\rm mult}/{\rm d}p_{\rm T}$ is the yield of electrons from heavy-flavour hadron decays in p--Pb collisions at $\sqrt{s_{\rm NN}}$~= 5.02~TeV measured in a given multiplicity class.

The ratio of the nuclear modification factor of electrons from heavy-flavour hadron decays in central multiplicities with respect to peripheral collisions, $\ensuremath{Q_{\rm cp}}\xspace$, was calculated as
 
\begin{equation}
Q_{\rm  cp} = \frac{\langle T_{\rm  pPb}^{\rm peripheral} \rangle}{\langle T_{ \rm  pPb}^{\rm central} \rangle}\frac{{\rm d}N^{ \rm  pPb}_{\rm central}/{\rm d}p_{ \rm T}}{{\rm d}N^{ \rm  pPb}_{\rm peripheral}/{\rm d}p_{\rm  T}},
\label{eq:Qcp}
\end{equation}     

where $ \langle T_{\rm pPb}^{\rm central} \rangle$ and $ \langle T_{\rm pPb}^{\rm peripheral} \rangle$ are the average nuclear overlap functions in the most central multiplicity interval and in the most peripheral multiplicity classes, respectively. The $ {\rm d}N^{\rm pPb}_{\rm central}/{\rm d}p_{\rm  T}$ is the yield of electrons from heavy-flavour hadron decays in the most central multiplicity interval and ${\rm d}N^{\rm pPb}_{\rm peripheral}/{\rm d}p_{\rm  T}$ is the corresponding yield in the most peripheral multiplicity class.

 The $Q_{\rm pPb}$ and $Q_{\rm cp}$ were measured within the \pt interval of 2 $< \pt<$ 16 \GeVc and the centrality ranges were selected as 0-20\%, 20-40\%, 40-60\%, and 60-100\%.  
 The measurements of electron production were performed in the midrapidity region in the centre-of-mass of the colliding system. 
 This corresponds to the asymmetric range -1.07 $< y_{\rm cms} <$ 0.14, since the centre-of-mass system moves with a rapidity of
$\Delta y_{\rm cms}$~= 0.465 in the direction of the proton beam, 
due to the different energies per nucleon of the proton and the lead beams.
The  \RpPb was measured in the high-\pt region (8 $< \pt <$ 20 \GeVc)
 updating the results for the momentum range 8-12 \GeVc and extending the \pt reach of the previously reported measurement~\cite{Adam:2015qda}.

The paper is organised as follows.  Section~2 describes the detector setup, data sample, and event selection criteria. Section~3 addresses the analysis details including the electron identification strategy. Systematic uncertainties are described in Sec.~4. Section~5 describes the pp reference. Section~6 presents the results. A summary is given in Sec.~7.

\section{Experimental apparatus, data sample, and event selection}\label{Event_selection}
\subsection{Experimental apparatus}
Detailed descriptions of the ALICE detectors can be found in ~\cite{Evans:2008zzb, Aamodt:2008zz, Abelev:2014ffa}. Electrons were reconstructed at midrapidity using the Inner Tracking System (ITS), the Time Projection Chamber (TPC), and the Electromagnetic Calorimeter (EMCal). The detectors are located inside a solenoidal magnet, which generates a magnetic field $B$ = 0.5 T along the beam direction. Event triggering was performed by the V0 detector, which consists of two scintillator arrays. The neutron Zero-Degree Calorimeters (ZNC) were used as a centrality estimator.   

The closest detector to the interaction point is the ITS~\cite{Aamodt:2010aa}, which is composed of six cylindrical layers of silicon detectors, located at radii between 3.9 cm and 43 cm. The two innermost layers form the Silicon Pixel Detector (SPD) which covers the pseudorapidity range $|\eta| <$ 2.0. The two intermediate layers form the Silicon Drift Detector (SDD) and the two outer layers consist of double-sided Silicon Strip Detector (SSD). Both cover a pseudorapidity range of $|\eta| <$ 0.9. The ITS can measure the charged-particle impact parameter (the distance of closest approach to the vertex) with a resolution better than 75 $\mu$m for transverse momenta $p_{\rm T} >$ 1~\GeVc~~\cite{Aamodt:2010aa}. It therefore has an important role in reconstructing the primary and secondary vertices. 

The main ALICE tracking device at midrapidity is the Time Projection Chamber ~\cite{Alme:2010ke}. It is a large cylindrical drift detector currently filled with a Ne-CO$_{2}$ gas mixture surrounding the ITS and extending from 85~cm to 247~cm in the radial direction and from -250 cm to +250 cm along the beam axis. The TPC covers $|\eta| <$ 0.9 and full azimuth for the maximum charged-particle track length of 159 reconstructed space points. The TPC enables charged-particle tracking beyond the ITS and particle identification via the measurement of the specific ionisation energy loss (d$E$/d$x$) with a resolution of up to 5.5\%~\cite{Acharya:2018hzf}.      

The EMCal~\cite{Cortese:2008zza} is a layered lead-scintillator sampling electromagnetic calorimeter. In Run-1 at the LHC, it covered 107$^{\circ}$ in azimuth and $|\eta| <$ 0.7 in pseudorapidity. The front face of the EMCal is situated about 450 cm from the beam axis in the radial direction. The 3072 modules are arranged in 10 full-sized and 2 one-third-sized  supermodules, consisting of 12 $\times$ 24 and 4 $\times$ 24 modules, respectively. The EMCal has 12288 towers, and each tower has a size of 6x6 cm$^{2}$. The energy resolution of the EMCal is $\sigma_{E}/E$ = 4.8\%/$E$ $\oplus$ 11.3\%/$\sqrt{E}$ $\oplus$ 1.7\%, where $E$ is the energy in GeV~\cite{Acharya:2018hzf}. 

The V0 detector~\cite{Abbas:2013taa} consists of two arrays of scintillator tiles at both forward, 2.8 $< \eta <$ 5.1 (V0A) and backward, -3.7 $< \eta <$ -1.7 (V0C) pseudorapidity regions. They are placed at distances $z$ = 3.4 m (V0A) and $z$ = -0.9 m (V0C) from the nominal interaction point and have full azimuthal coverage. This detector was used for triggering, and event centrality determination. The Zero Degree Calorimeters (ZDCs)~\cite{Oppedisano:2009zz}, located close to the beam pipe, measure the spectator protons and neutrons. They consist of two sets of neutron (ZNA and ZNC) and proton (ZPA and ZPC) calorimeters positioned on either side of the interaction point at $z$ = $\pm$112.5 m. They are used to remove the contamination from beam-background interactions and also to determine the centrality of the collisions.

\subsection{Data sample and event selection}
This analysis used 100 million minimum-bias (MB) events and 0.9 million events triggered by a high energy deposit in the EMCal, both recorded during the  {\mbox{p--Pb}} run in 2013. The MB trigger requires a coincidence of signals in the V0A and V0C detectors. The MB dataset was used for the measurement of electrons from heavy-flavour hadron decays in the range 2 $<\pt<$ 8~\GeVc. The EMCal trigger was used to record electrons at high-\pt and therefore extends the kinematic reach of the MB measurements. In this analysis, the data were collected with a Level-1 trigger \cite{Bourrion:2010js, Bourrion:2012vn}, which is a hardware trigger consisting of the sum of energy in a sliding window of 4x4 towers above a given threshold, where a tower is the smallest segmentation of the EMCal. The {\mbox{p--Pb}} data collected with the EMCal trigger with energy thresholds of about 7 GeV and 11~GeV were used to measure charged particle tracks in the ranges 8 $<\pt<$ 12~\GeVc and 12 $<\pt<$ 20~\GeVc, respectively.

The primary vertex was reconstructed using tracks in the ITS and TPC. A selection on the vertex position along the beam axis ($z$) within $\pm$10 cm from the nominal interaction point was applied in the analysis.

The integrated luminosity analysed was \mbox{$L_{\mathrm{int}} = 47.8 \pm 1.6~\mu\mathrm{b}^{-1}$} for MB data, and \mbox{$L_{\mathrm{int}} = 0.191 \pm 0.018 ~\mathrm{nb}^{-1}$} (\mbox{$L_{\mathrm{int}} = 1.62 \pm 0.15~ \mathrm{nb}^{-1}$}) for the lower (higher) EMCal trigger threshold.

\subsubsection{Centrality determination}

The centrality estimation was based on the ZNA detector which measures the multiplicity of neutrons produced in the interaction. The event properties (the number of participant nucleons, $N_{\rm part}$, and the number of binary collisions, $N_{\rm coll}$) were calculated based on a Glauber model coupled to a negative binomial distribution, as described in~\cite{ALICE-PUBLIC-2018-003}.
Due to its large $\eta$-separation from the central barrel detectors, the ZNA is expected to be the least biased centrality estimator, as demonstrated in ~\cite{Adam:2014qja}.
The values of $N_{\rm part}$, $N_{\rm coll}$, and the nuclear overlap function $T_{\rm pPb}$ were obtained using the hybrid method. 

The hybrid method relies on two main assumptions: the first is to assume that an event selection
based on ZNA does not introduce any bias on the bulk at midrapidity and on high$-\pt$ particle production; the second assumption is that
the $N_{\rm coll}$ determination is based on a particular scaling  for  particle  multiplicity, where it is assumed  that  the  charged-particle multiplicity measured at midrapidity scales with the number of participants~\cite{Adam:2014qja, ALICE-PUBLIC-2018-011}.

The values of the average nuclear overlap function $\langle T_{\rm pPb} \rangle$ obtained with the ZNA in the four multiplicity classes used for the analysis were 
obtained using the formula $\langle T_{\rm pPb}^{\rm mult}  \rangle$ = $ \langle N_{\rm coll}^{\rm mult} \rangle_{i} / \sigma_{\rm NN}$, where $N_{\rm coll}^{\rm mult}$ is the number of binary collisions calculated in each multiplicity interval and $\sigma_{\rm NN}$ = (67.6 $\pm$ 0.6) mb is the inelastic nucleon--nucleon cross section at $\sqrt{s_{\rm NN}}$ = 5.02 TeV, estimated from interpolating data at different centre of mass energies~\cite{ALICE-PUBLIC-2018-011}. The values of $ \langle T_{\rm pPb}  \rangle$ are reported in Tab.~\ref{TableTpPb}. 

\begin{table}
\begin{center}
\begin{tabular}{l l l}
\hline
Centrality class & $\langle T_{\rm pPb}  \rangle$\\ 
\hline
0-20\% & 0.1649 $\pm$ 5.4\% \\
\hline 
20-40\% & 0.1374 $\pm$ 2.4\% \\
\hline
40-60\% & 0.1016 $\pm$ 5.1\% \\
\hline
60-100\% & 0.0459 $\pm$ 5.2\% \\
\hline
\end{tabular}
\caption{$\langle T_{\rm pPb}  \rangle$ values in p--Pb collisions at $\sqrt{s_{\rm NN}}$ = 5.02 TeV obtained with the hybrid method using the ZNA, as described in ~\cite{ALICE-PUBLIC-2018-011}.} 
\label{TableTpPb}
\end{center}
\end{table}

\subsubsection{Trigger scaling factor}

\begin{figure}[h!]
\centering
\includegraphics[width=0.6\linewidth]{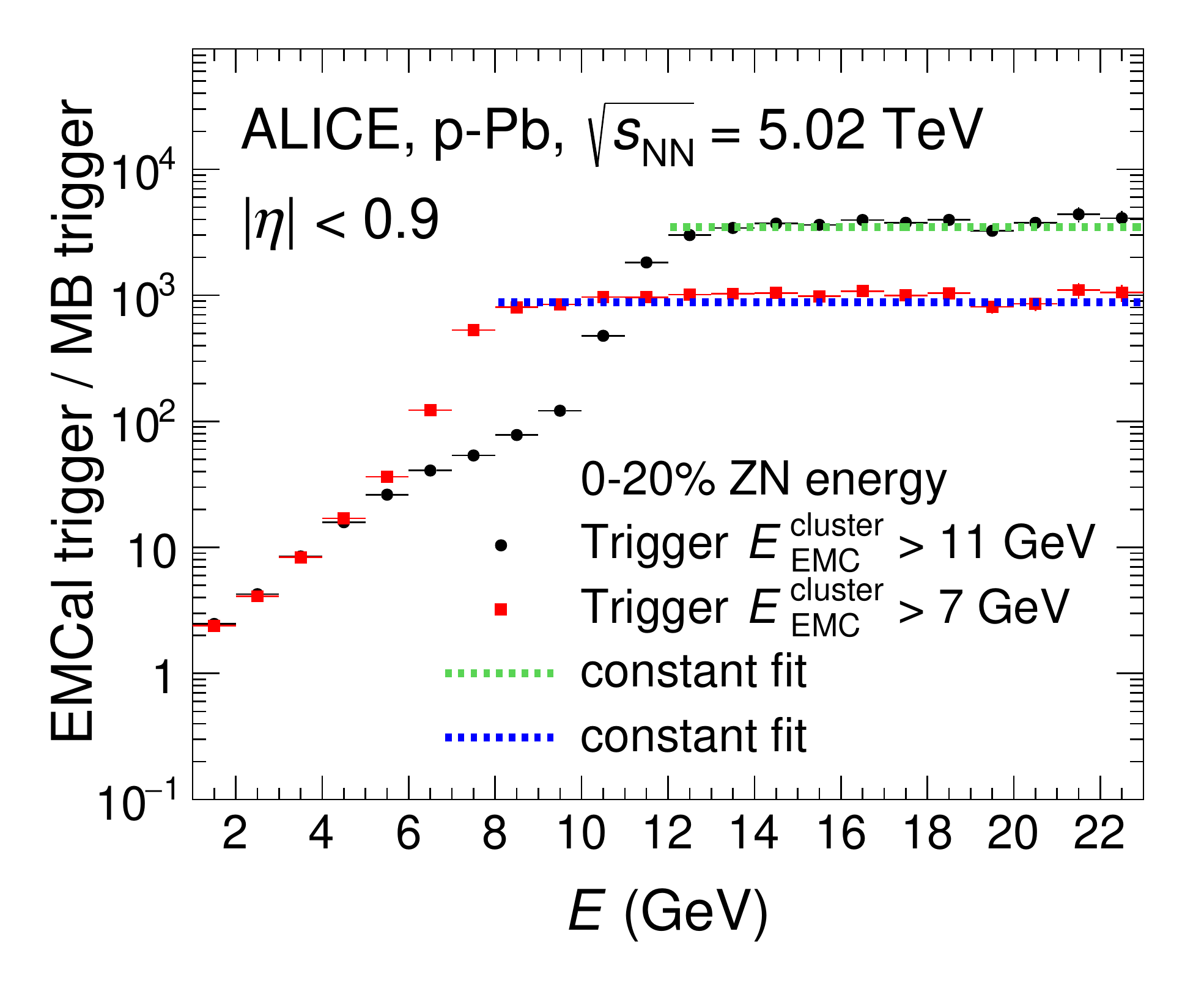}
\caption{Example of a trigger turn-on curve for the multiplicity class 0-20\%. Each scaling factor was obtained by fitting a constant to the plateau region (dashed lines) of the distribution. The resulting values are summarised in Tab.~\ref{tab:rej_factor}.} 
\label{fig:RF}
\end{figure}

Due to the trigger enhancement of electrons at high-\pt, the yields obtained using the EMCal triggered data samples were corrected by the trigger scaling factor in each centrality class.
This correction was obtained via a data-driven method where the cluster energy distribution in triggered-data was divided by the cluster energy distribution in minimum-bias triggered data. The ratio of these distributions give the turn-on curve. Figure~\ref{fig:RF} shows one example of the turn-on curve ($E_{\rm EMC}^{\rm cluster} > 11$ GeV and $E_{\rm EMC}^{\rm cluster} > 7$ GeV) of the trigger for the centrality class 0-20\%, as a function of the energy for all clusters in the EMCal detector.

The scaling factor was obtained by fitting a constant to the plateau of the turn-on curve in an interval above the trigger threshold where the distribution flattens. The values obtained for the scaling factor are summarised in Tab.~\ref{tab:rej_factor}. The uncertainties on the fits are approximately 1\% and the systematic uncertainties were obtained using different fit ranges on the plateau (as discussed in Sec.~\ref{systematic}).

\begin{table}[h!]
\centering
\begin{tabular}{lllll}
\cline{1-3}
Centrality class & Scaling factor for $E_{\rm EMC}^{\rm cluster} > 11 $ GeV  & Scaling factor for $E_{\rm EMC}^{\rm cluster} > 7 $ GeV &  &  \\ \cline{1-3}
0-20\%    & 3348 $\pm$ 285  & 873 $\pm$ 79  &  &  \\
20-40\%   & 4070 $\pm$ 346  & 1078 $\pm$ 97  &  &  \\
40-60\%   & 5400 $\pm$ 459   & 1484 $\pm$ 134 &  &  \\
60-100\%  & 11113 $\pm$ 945 & 3161 $\pm$ 284  &  &  \\
0-100\%   & 5439 $\pm$ 462  & 1432 $\pm$ 129 &  &  \\ \cline{1-3}
\end{tabular}
\caption{Values of the EMCal trigger scaling factor and their systematic uncertainties for the $E_{\rm EMC}^{\rm cluster} > 11$ GeV trigger and the $E_{\rm EMC}^{\rm cluster} > 7$ GeV trigger.}
\label{tab:rej_factor}
\end{table}

\section{Analysis}\label{analysis}

The electron identification (eID) was performed using a combination of two different strategies.
For the low-\pt interval ($2 < \pt < 8$  \GeVc) only the TPC signal was used to identify electrons, since in this \pt range the specific ionisation energy loss (\dEdx) of the electrons in the TPC is well separated from that of the hadrons. For the high-\pt ($8 < \pt < 16$  \GeVc for the multiplicity analysis and $8 < \pt < 20$ \GeVc for the integrated analysis) measurements, the combination of both the TPC and the EMCal detectors was used, since above 8 \GeVc the \dEdx distribution of pions begins to merge with the \dEdx distribution of electrons. The usage of the EMCal reduces the amount of hadron contamination, since they can be well separated using the ratio of energy ($E$) deposited in the EMCal to the momentum ($p$) of the tracks. For electrons, $E/p$ is around unity since they deposit all of their energy in the EMCal and their mass is relatively small compared to their energy. Therefore, $E/p$ can be used to select electrons and reject hadrons.

The charged-particle track selection criteria used in this analysis are similar to that used in previous measurements of electrons from heavy-flavour hadron decays in pp collisions ~\cite{Abelev:2014gla, Abelev:2012xe} and \pPb collisions ~\cite{Adam:2015qda}.
For the track quality selection, a minimum of 100 clusters in the TPC were required and at least 4 (3) clusters in the ITS for the MB (EMCal trigger) data sample. The requirement of two SPD hits reduces the number of electrons from $\gamma$ conversions in the detector material. Within the EMCal acceptance there are dead regions in the first layer of the SPD, therefore at high \pt (\pt $> 8$ \GeVc) only one hit was required.
The tracks used for the analysis were also required to be close to the primary vertex. The distance of closest approach (DCA) to the primary vertex was required to be $\rm{DCA}_{xy} < 2.4$ cm in the transverse plane and $\rm{DCA}_{z} < 3.6$ cm in the longitudinal direction (beam axis) in order to reject background  and non-primary tracks.

\begin{figure}[h!]
\centering
\includegraphics[width=0.5\linewidth]{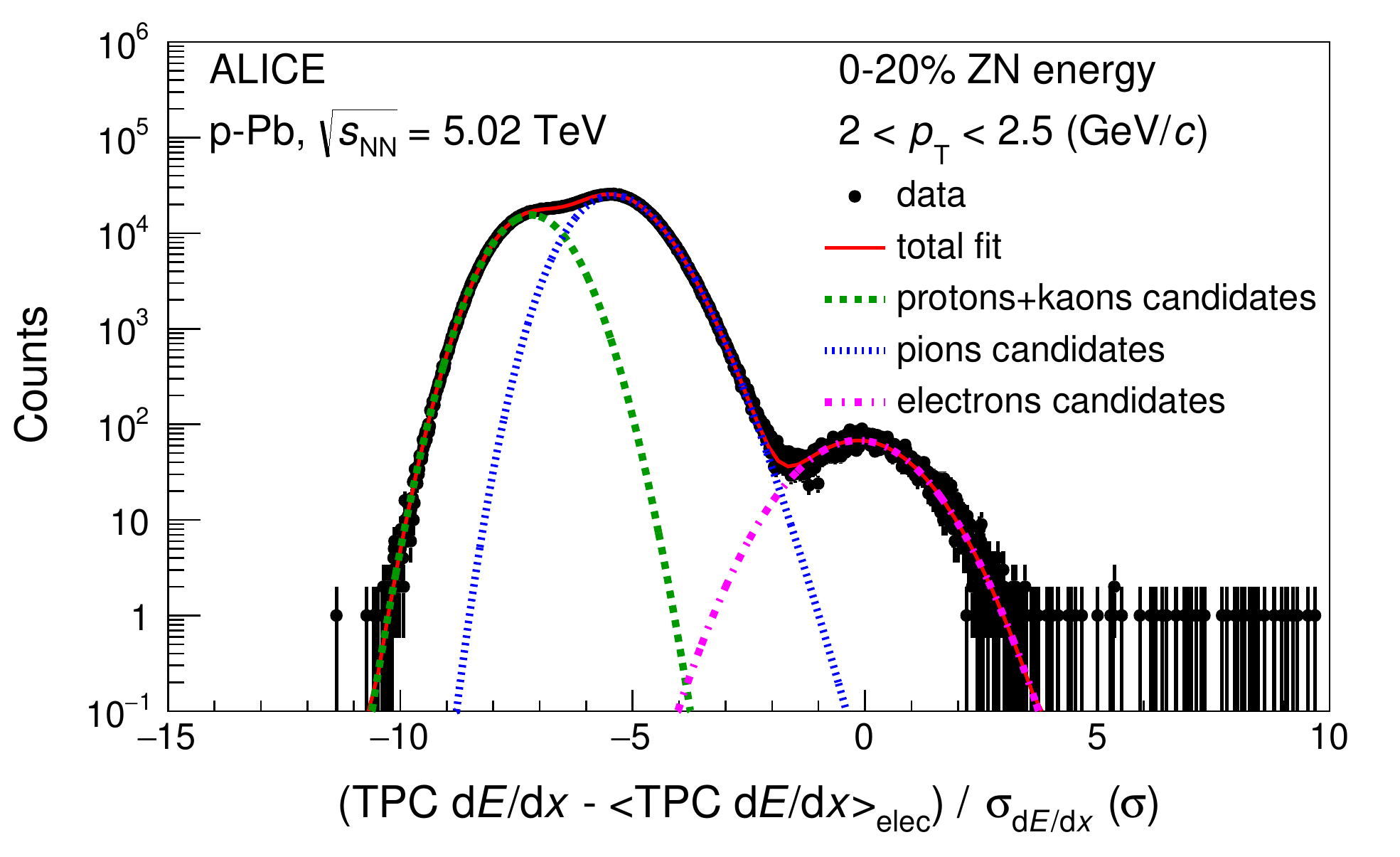}
\caption{The measured d$E$/d$x$ in the TPC expressed as a standard deviation from the expected energy loss of electrons, normalised by the energy-loss resolution ($\sigma_{\rm TPC}$) 
for $2 < \pt < 2.5$ \GeVc. The various curves are the different fit function results for the different peaks of the distribution. A Gaussian distribution, centered around zero, describes the electron candidates, and the pions and protons are the curves around $n_{\sigma}^{\rm TPC} < -4$ (Landau distribution multiplied by an Exponential distribution) and $n_{\sigma}^{\rm TPC} < -8$ (Gaussian distribution), respectively.}
\label{fig:TPCnsigma}
\end{figure}

After  selecting high quality tracks, the energy loss in the TPC was used to select electron candidates. The selection was based on the number of standard deviations of the measured signal from the signal expected if the track was an electron, $n_{\sigma}^{\rm TPC}$.
An example of the $n_{\sigma}^{\rm TPC}$ distribution is shown in Fig.~\ref{fig:TPCnsigma} for $2 < \pt < 2.5$ \GeVc. 
A Gaussian distribution, centered around zero, describes the electron candidates, and the pions and protons are the curves around $n_{\sigma}^{\rm TPC} = -4$ and $n_{\sigma}^{\rm TPC} = -8$, respectively,  for this $\pt$ bin. Pions are described by a Landau distribution multiplied by an exponential distribution, while the protons are described by a Gaussian distribution.
For the low $\pt$ ($2 < \pt < 8$ \GeVc) analysis, electrons were selected by requiring $0< n_{\sigma}^{\rm TPC} < 3$ to avoid an overlap with the pion band. For this selection, the hadron contamination is negligible for $2< \pt < 6$ \GeVc and 0.5\% for $6< \pt < 8$ \GeVc.

\begin{figure}[h!]
\centering
\includegraphics[width=0.95\linewidth]{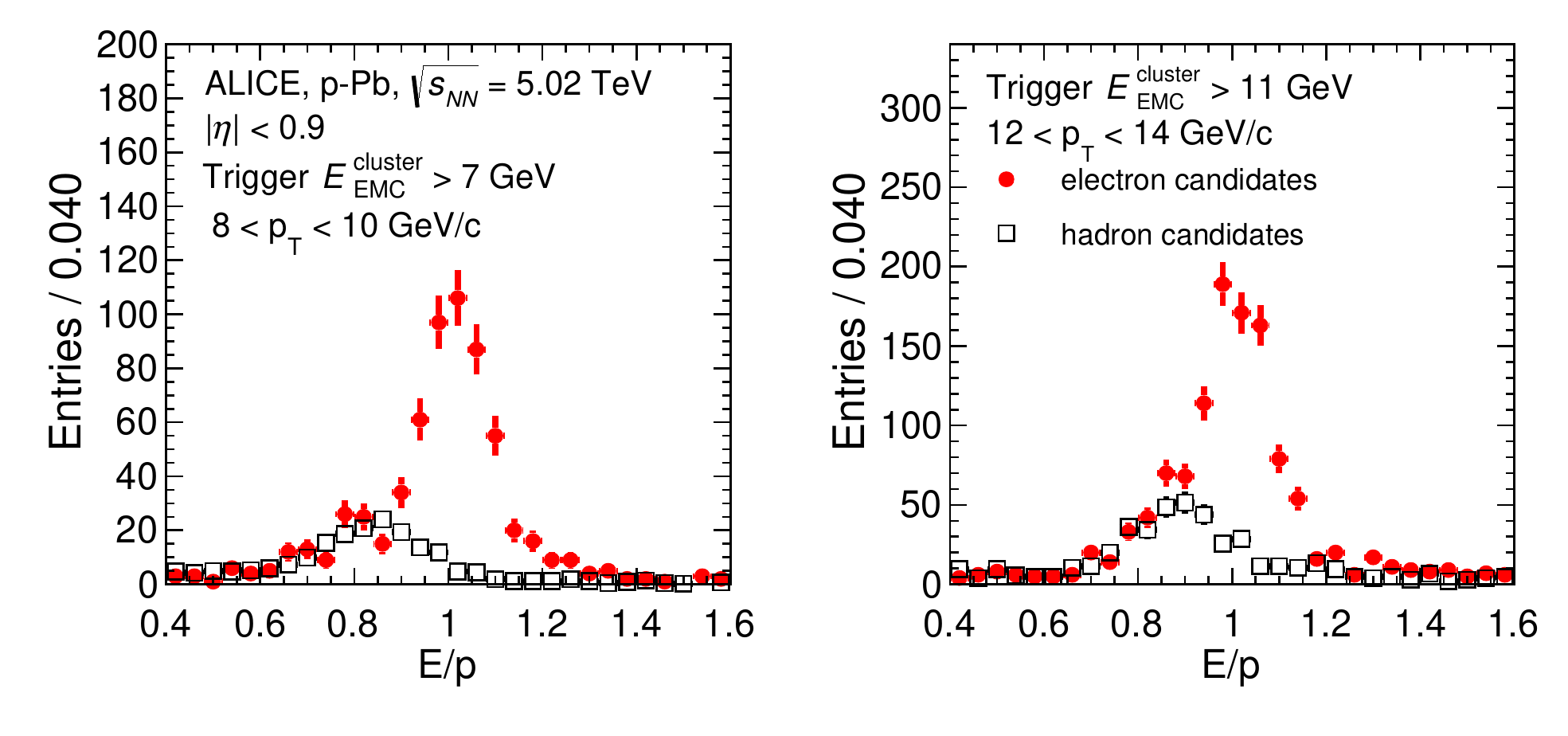}
\caption{$E/p$ distribution for integrated centrality for $8 < \pt < 10$ \GeVc for the lower EMCal threshold triggered events (left) and  for $12 < \pt < 14$ \GeVc
 for the higher EMCal threshold triggered events (right). The distributions are shown for electron candidates selected by the TPC ($-1 <n_{\sigma}^{\rm TPC} < 3$) (solid symbols) and for hadron candidates (open symbols) selected by the TPC $n_{\sigma}^{\rm TPC} < -3.5$. }
\label{fig:EoverP}
\end{figure}

For the high-$\pt$ ($8 < \pt < 16$  \GeVc for the multiplicity analysis and $8 < \pt < 20$ \GeVc for the integrated analysis) measurements, where the EMCal trigger was used, the electron candidates were selected in the band $-1 < n_{\sigma}^{\rm TPC} < 3$ and  $E/p$ distributions were used to remove the hadron contamination and to count the electron candidates.    
Figure~\ref{fig:EoverP} shows the $E/p$ distribution for $8 < \pt < 10$ \GeVc for the lower EMCal trigger threshold (left) and for $12 < \pt < 14$ \GeVc for the higher EMCal trigger threshold (right) after requiring $-1 < n_{\sigma}^{\rm TPC} < 3$. Electrons are expected to be around unity while a hadron peak arises around $E_{\rm th}$/\pt, where $E_{\rm th}$ is the EMCal trigger threshold.

To decrease the amount of hadron contamination, a condition on the electromagnetic shower shape was used~\cite{Adam:2016khe, Abelev:2014ffa}. 
The shower shape produced in the calorimeter has an elliptical shape which can be characterised by its two axes: $\sigma_{\rm long}^{2}$ for the long axis  and $\sigma_{\rm short}^{2}$ for the short axis. A rather lose selection of $\sigma_{\rm short}^{2} < 0.3$ was chosen, since it reduces the hadron contamination while at the same time does not significantly affect the electron signal.
The hadron contamination was estimated in each multiplicity interval by measuring $E/p$ for hadrons, after requiring $n_{\sigma}^{\rm TPC} < -3.5$. The $E/p$ distribution for hadrons was scaled to match the electron's $E/p$ distribution in the range $0.4 < E/p < 0.7$. The electron yield was obtained by integrating the distribution for $0.8 < E/p < 1.2$ and subtracting the hadronic contribution statistically. For $8 < \pt < 10$ \GeVc the hadron contamination is around 18\% and for $12 < \pt < 14$ \GeVc  it is around 35\% for integrated centrality.

The background electrons, which are mainly from electrons produced by $\gamma$ conversions ($\gamma \rightarrow  e^{+}e^{-}$) in the material and Dalitz decays of neutral mesons, e.g. $\pi^0 \rightarrow \gamma\ \rm{e^+} \rm{e^-}$ and $\eta \rightarrow \gamma\ \rm{e^+   e^-}$, were removed using an invariant mass method~\cite{Adam:2015qda}. Since these electrons are produced in $\rm{e^+ e^-}$ pairs and therefore their invariant mass should be less than the pion mass, a selection of 0.15 GeV/${c^2}$ was required. The efficiency was determined using two Monte Carlo (MC) samples, where, in both of them, pp collisions generated with PYTHIA were embedded in \pPb events simulated by HIJING~\cite{Gyulassy:1994ew}.  
The first sample was generated requiring that each PYTHIA event contains a $\rm c\overline{c}$ or $\rm b\overline{b}$ pair decaying semileptonically, using the generator PYTHIA v6.4.21~\cite{Sjostrand:2006za} with the \mbox{Perugia-0} tune~\cite{Skands:2010ak}. This enhancement of heavy-flavour electrons increases the statistical precision of the total electron efficiency (reconstruction and identification efficiency) determination at intermediate and high \pt.
The second sample used in this analysis included an enhancement of $\pi^{0}$ and $\eta$ mesons in order to increase the statistical precision of the efficiency of finding pairs using the invariant mass method. The simulated $\pi^{0}$ and $\eta$ \pt distributions were reweighted to match the measured shapes. The $\pi^{0}$ spectra were estimated as the average of the spectra of $\pi^{+}$ and $\pi^{-}$~\cite{Abelev:2013haa, Adam:2016dau}  and the $\eta$ spectra were estimated using $m_{\rm T}$ scaling, as in ~\cite{Adam:2015qda}. 
The efficiencies were around 70\% for the low \pt ($2 < \pt < 8$~\GeVc) analysis and around 85\% for the high \pt bins ($8 < \pt < 16$~\GeVc), independent of the multiplicity class.

The $\pt$-differential invariant cross section $\sigma_{\rm hfe}$ of electrons from heavy-flavour hadron decays (hfe) was calculated as
\begin{equation}
\frac{1}{2\pi\pt}\frac{{\rm d}^{2}\sigma_{\rm hfe}}{{\rm d}\pt{\rm d}y}=
\frac{1}{2}
\frac{1}{2\pi \pt^{\rm{centre}}} 
\frac{1}{\Delta y \Delta \pt}
\frac{ N_{\rm hfe}^{\rm raw}}{(\epsilon^{\rm geo} \times \epsilon^{\rm reco} \times \epsilon^{\rm eID})}
\frac{\smbeq}{N},
\label{eq:cross_section}
\end{equation}

where $\pt^{\rm{centre}}$ is the centre of the \pt\ bin, $\Delta\pt$ is the width of the \pt\ bin, and $\Delta y$ is the rapidity range where the analysis was performed. $N$ is the number of events analysed and $\smb = 2.09\pm0.07$~b is the \pPb cross section for the minimum-bias V0 trigger condition~\cite{Abelev:2014epa}. In the case of the analysis using the EMCal trigger, $N$ is the number of events that satisfy the trigger requirements multiplied by the trigger scaling factor. 
$ \epsilon^{\rm reco}$ is the track reconstruction efficiency,  $\epsilon^{\rm eID}$ is the electron identification efficiency, and $\epsilon^{\rm geo}$ is the acceptance of the detectors. $N_{\rm hfe}^{\rm raw}$ is the number of electrons from heavy-flavour hadron decays, obtained by subtracting the background electrons from the inclusive electron distributions.

For the MB data, the total efficiency including acceptance is around 28\% and for the EMCal triggered data, due to its finite acceptance, the value is around 12\%, independent of multiplicity class in the measured \pt range.
To take into account the momentum resolution and the energy loss due to bremsstrahlung in the detector material, an unfolding procedure based on Bayes' theorem was applied ~\cite{DAgostini:1999gfj, Grosse-Oetringhaus:2009mka}.
The remaining residual background  originating from semileptonic kaon decays, dielectron decays of J/$\psi$ mesons, and W boson decays to electrons was evaluated using simulations and were removed from the electron yield. While the contribution from kaon decays is negligible, J/$\psi$ mesons have a maximum contribution of 2.9\% around 3.5 \GeVc and W boson decays have a maximum contribution of 2.5\% at 20 \GeVc.

\section{Systematic uncertainties}\label{systematic}

Systematic uncertainties were estimated as a function of \pt by repeating the analysis and varying the selection criteria in each centrality class. 
 For the \RpPb and $Q_{\rm pPb}$ measurements, the uncertainties were evaluated by analysing the invariant yield separately for each centrality class. 
For the $Q_{\rm cp}$ measurement, the systematic uncertainties were estimated by evaluating the variations directly on the $Q_{\rm cp}$ for each centrality interval. 
The different sources of systematic uncertainties are further discussed in this section.

The systematic uncertainties on the track selection, track matching, and electron identification were obtained via multiple variations of the selection criteria. For the track selection the minimum number of space points in the TPC and the hits in the ITS were varied. The systematic uncertainty for the matching between the ITS and TPC was taken as 3\% according to  ~\cite{Abelev:2014dsa}. The TPC and EMCal track matching uncertainty was assigned to be 1\%, as determined by varying the size of the matching window in pseudorapidity and azimuth for electron candidates that were extrapolated to the calorimeter.
The restriction on $n_{\sigma}^{\rm TPC}$ was varied to determine the systematic uncertainty on electron identification with the TPC. For the EMCal based electron identification, the $E/p$ range and shower shape criteria were varied around their nominal value. 

The uncertainties on the measurement of the background were obtained by varying the invariant mass criteria of the electron-positron pairs, 
the minimum \pt of the tracks paired with electron candidates, 
and the opening angles between the electron-positron pairs. 
The uncertainty from the re-weighting procedure performed on the $\pi^{0}$ and $\eta$-meson \pt distributions in MC simulations was estimated by changing the weights by $\pm$10\% and for both a negligible effect on the yield measurement was found. The systematic uncertainties of the heavy-flavour electron yield due to the subtraction of the remaining background originating from semileptonic kaon decays and dielectron decays from J/$\psi$ mesons are negligible ($\sim$~0.06\%). This was estimated by changing the electron yields from the J/$\psi$ and kaon decays by $\pm$50\% and $\pm$100\%, respectively. The systematic uncertainty for the yield of electrons from W boson decays is also negligible ($<$ 0.5\%). It was measured by varying the yield of electrons from W boson decays by $\pm$15\%.    

For part of the analysed \pPb dataset, fewer high-\pt particles were observed for negative $\eta$ than positive $\eta$. The difference is related to distortions on the negative $\eta$ side of the TPC, and the effect was corrected using a data-driven method. The spectra of charged particles were obtained in both negative and positive  $\eta$ sides and the negative side was corrected in order to match the positive side. A systematic uncertainty of 5\% was assigned to cover remaining differences.

The systematic uncertainty for the EMCal trigger correction was obtained by changing the fit ranges on the plateau of the turn-on curve. There is a 8.5\% deviation for the highest threshold and 9\% for the lowest threshold, which is assigned as the systematic uncertainty. It is centrality and \pt independent and applied to the yield obtained using the triggered data. 
 
The systematic uncertainties are summarised in Tab.~\ref{Tablesys1}.
Since the sources are uncorrelated, they were added in quadrature to give a total systematic uncertainty, which is 6\% for MB data and 13\% (12\%) for EMCal lower threshold (higher threshold) triggered data. For the $Q_{\rm cp}$ measurement, they are 5\% and 10\%, respectively. 
In the table the systematic uncertainties are presented in the \pt range of 2 $<$  \pt $<$ 8 \GeVc (TPC only yield$_{\rm cent} $/$Q_{\rm cp}$), EMCal triggered analysis (TPC+EMCal  yield$_{\rm cent}$/$Q_{\rm cp}$) for the \pt range of 8  $< $ \pt $<$ 16 \GeVc and EMCal triggered analysis (for integrated centrality,  TPC+EMCal  yield$_{\rm int}$), for the \pt range of 8 $< $ \pt $<$ 20 \GeVc.   

\begin{table}
\begin{center}
\begin{tabular}{l l l l }
\hline
Sources & TPC only (yield$_{\rm cent}$ / $Q_{\rm cp}$) (\%)& TPC+EMCal (yield$_{\rm int}$ / yield$_{\rm cent}$ / $Q_{\rm cp}$) (\%) \\
\hline 
\hline
Track selection & 2 / 2 & No effect \\
\hline
ITS-TPC matching & 3 / cancels & 3 / 3 / cancels \\
\hline
TPC-EMCal matching & not applicable & 1 / 1 / cancels \\
\hline
TPC eID & 3 / 3 & 5 / 5 / 5 \\
\hline
EMCal eID & not applicable & 3 / 3 / 3 \\
\hline
Invariant mass method & 3 / 3 & 3 / 3 / 3 \\
\hline
J/$\psi$ electron background & negl. &  0.06 / 0.06 / cancels \\
\hline
W electron background & negl.  &  0.3 / 0.3 / cancels \\
\hline
$\pi^{0}$, $\eta$ weight & negl. &  negl. \\
\hline
$\eta$ A vs C side & not applicable & 5 / 5 / 7 \\
\hline
EMCal trigger correction & not applicable & 8.5 and 9  / 8.5 and 9 / cancels \\
\hline
\hline
Total &  6 / 5  & 12  and 13   / 12  and 13 / 10\\
\hline
\end{tabular}
\caption{Systematic uncertainties for the TPC only and TPC+EMCal analysis in percentual values. yield$_{\rm int}$ and yield$_{\rm cent}$ represent the invariant yield for integrated centrality and for different centrality classes, respectively. For the EMCal trigger correction, the two values presented are for $E_{\rm EMC}^{\rm cluster} > 11$ GeV and $E_{\rm EMC}^{\rm cluster} > 7$ GeV, respectively.}
\label{Tablesys1}
\end{center}
\end{table}

\section{pp reference}\label{ppref}

To measure the nuclear modification factor (\RpPb or $Q_{\rm pPb}$) a reference cross section for pp collisions at the same centre-of-mass energy is needed. 
The \RpPb results from~\cite{Adam:2015qda} are updated for $0.5 < \pt < 10$ \GeVc using a recent measurement of electrons from heavy-flavour hadron decays in pp collisions at $\sqrt{s}=5.02$ TeV~\cite{Acharya:2019mom}. Using the new pp reference, the \RpPb uncertainties are improved by a factor of 2-4, depending on the transverse momentum.

 For the higher \pt interval 10 $< \pt <$ 20 \GeVc , a scaling was performed using the ATLAS data~\cite{Aad:2011rr} at $\sqrt{s}$ = 7 TeV within the same \pt region. Since perturbative quantum chromodynamics (pQCD) calculations at fixed order with next-to-leading-log (FONLL) calculations~\cite{Cacciari:1998it, Cacciari:2001td, Cacciari:2012ny} describe the data at 5.02 TeV and 7 TeV within experimental and theoretical uncertainties, they were used  to scale the ATLAS data to 5.02 TeV.  
  The scaling is \pt dependent and based on the ratio of spectra at  7 TeV and 5.02 TeV.
Since the rapidity coverage of the ATLAS measurement is different ($|y| <$ 2 excluding 1.37 $< |y| <$~1.52) from this measurement ($|y| <$~0.6) the ratio of \pt-differential cross sections of heavy-flavour decay electrons measured in two different rapidity regions were corrected based on FONLL calculations.
The systematic uncertainties on the scaled ATLAS pp spectrum at $\sqrt{s}$ = 5.02 TeV range from 18\% to 13\% in the \pt bins used in this analysis. The statistical uncertainties are from the ATLAS measurement.     
   
In summary, in this paper,  \RpPb and $Q_{\rm pPb}$ are calculated using the pp reference measured by ALICE at $\sqrt{s}=5.02$ TeV~\cite{Acharya:2019mom} up to 10 \GeVc and using ATLAS data~\cite{Aad:2011rr} scaled to  5.02 TeV for $\pt > 10$ \GeVc.

\section{Results}\label{results}

\begin{figure}[h!]
\centering
\includegraphics[width=0.9\linewidth]{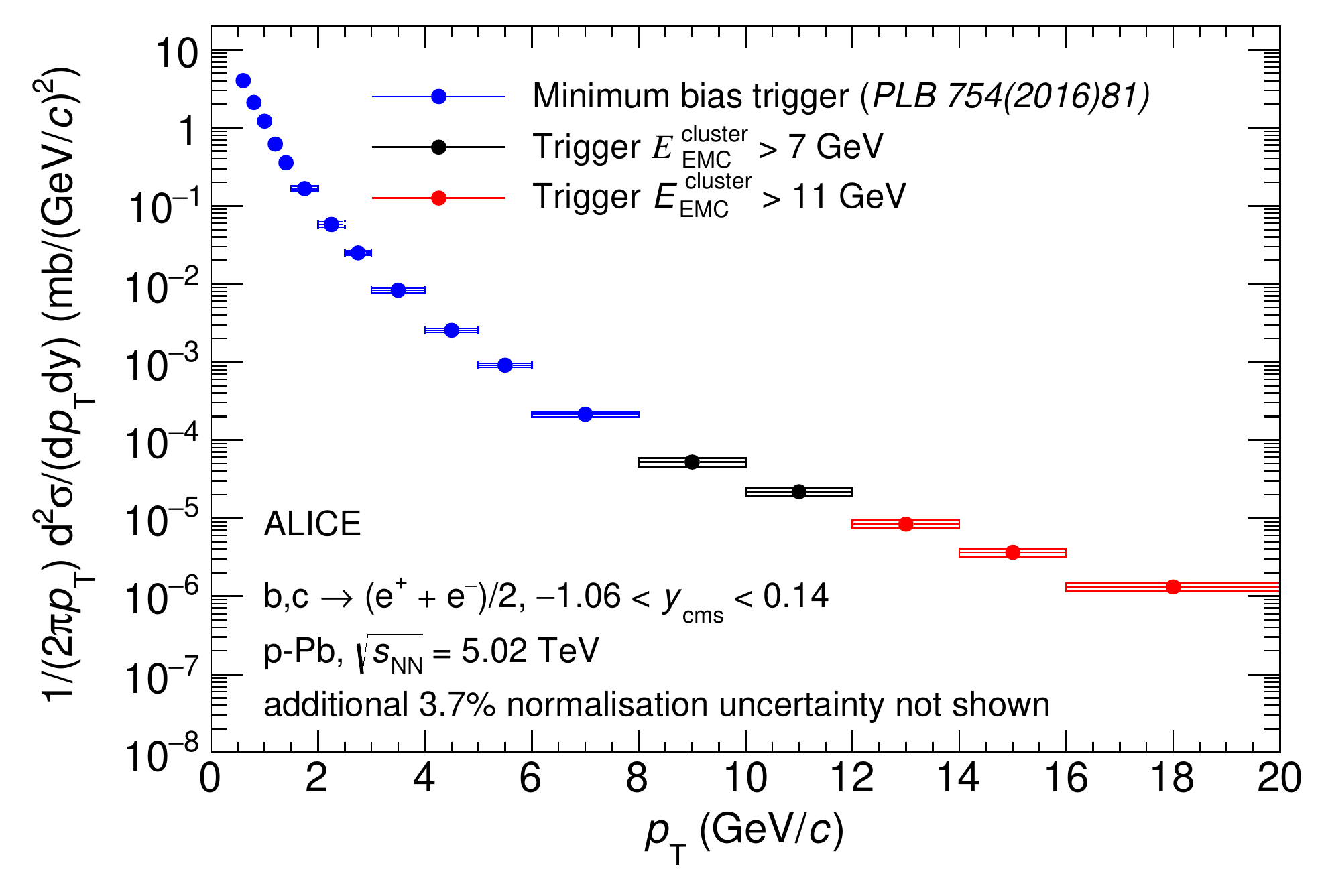}
\caption{The \pt-differential invariant cross section of electrons from heavy-flavour hadron decays  in p--Pb collisions at $\sqrt{s_{\rm NN}}$ = 5.02 TeV. The statistical uncertainties are indicated for both spectra by error bars and the systematic uncertainties are shown as boxes. The published result is shown for 0.5 $<$ \pt $<$ 8 \GeVc~\cite{Adam:2015qda}, and the measurement using the EMCal trigger is shown up to \pt = 20 \GeVc.}
\label{fig:cross_section1}
\end{figure}

The \pt-differential invariant cross section of electrons from semi-leptonic decays of heavy-flavour hadrons in p--Pb collisions at $\sqrt{s_{\rm NN}}$ = 5.02 TeV is shown in Fig.~\ref{fig:cross_section1} as a function of \pt. The published data which were measured using the TPC, TOF, and EMCal detectors~\cite{Adam:2015qda} are also shown in Fig.~\ref{fig:cross_section1}.
 In this work, the \pt-differential invariant cross section results are improved in the \pt range 8-12 \GeVc and extended up to \pt = 20 \GeVc, using the statistics collected with the EMCal trigger.

\begin{figure}[h!]
\centering
\includegraphics[width=0.9\linewidth]{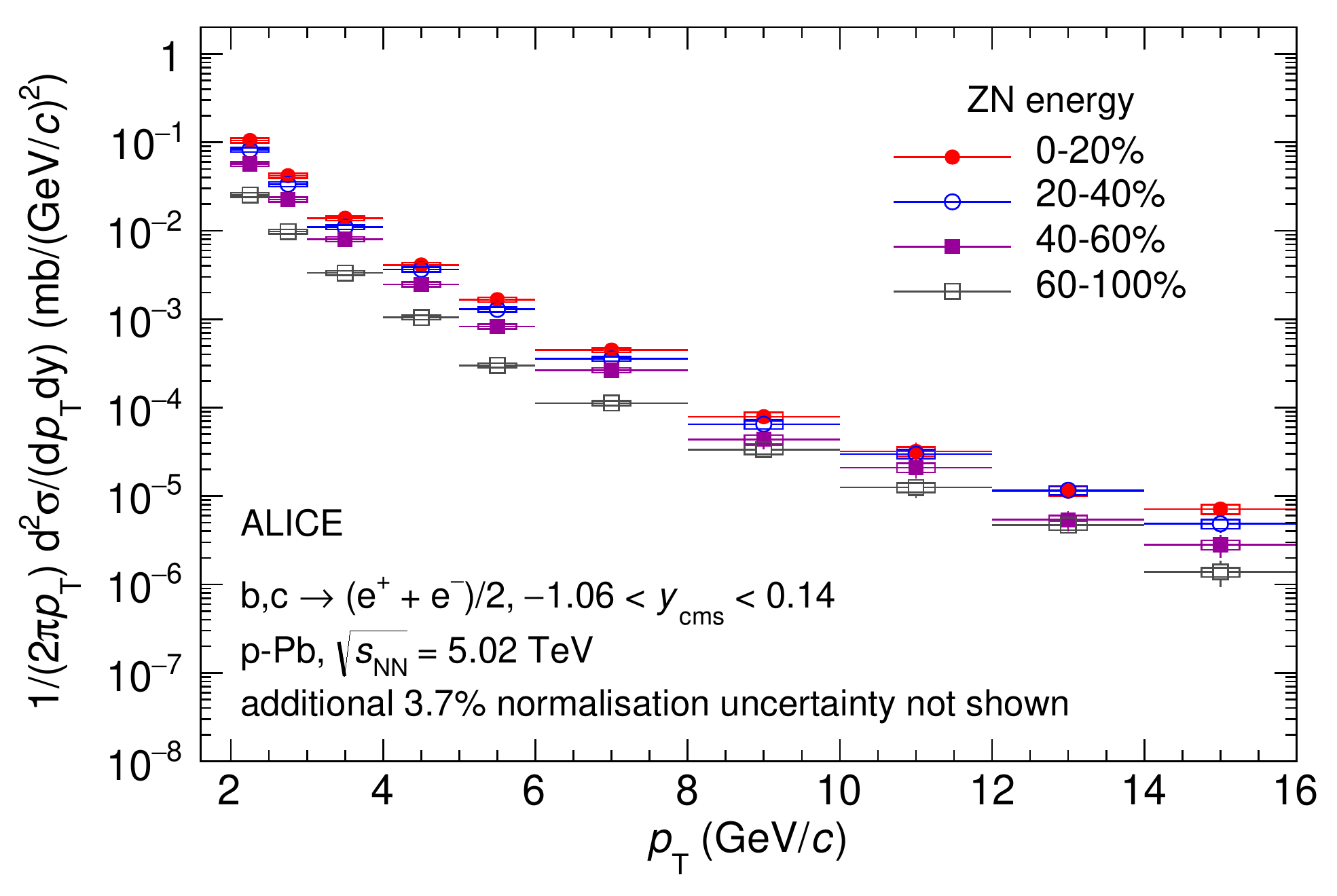}
\caption{The \pt-differential invariant cross section of electrons from heavy-flavour hadron decays in several charged-particle multiplicity classes in p--Pb collisions at $\sqrt{s_{\rm NN}}$ = 5.02 TeV. The statistical uncertainty of each spectrum is indicated by error bars and the systematic uncertainties are indicated by boxes.}
\label{fig:cross_section}
\end{figure}

 Figure~\ref{fig:cross_section} shows the cross section of electrons from semi-leptonic decays of heavy-flavour hadrons in p--Pb collisions at $\sqrt{s_{\rm NN}}$ = 5.02 TeV measured in different multiplicity classes and corrected for detector acceptance and efficiency.
The multiplicity classes were estimated based on the ZNA detector, as described in Sec.~\ref{Event_selection}, and the cross section of electrons from semi-leptonic decays of heavy-flavour hadrons were measured in 0-20\%, 20-40\%, 40-60\%, and 60-100\% multiplicity classes.

\begin{figure}[h!]
\centering
\includegraphics[width=0.9\linewidth]{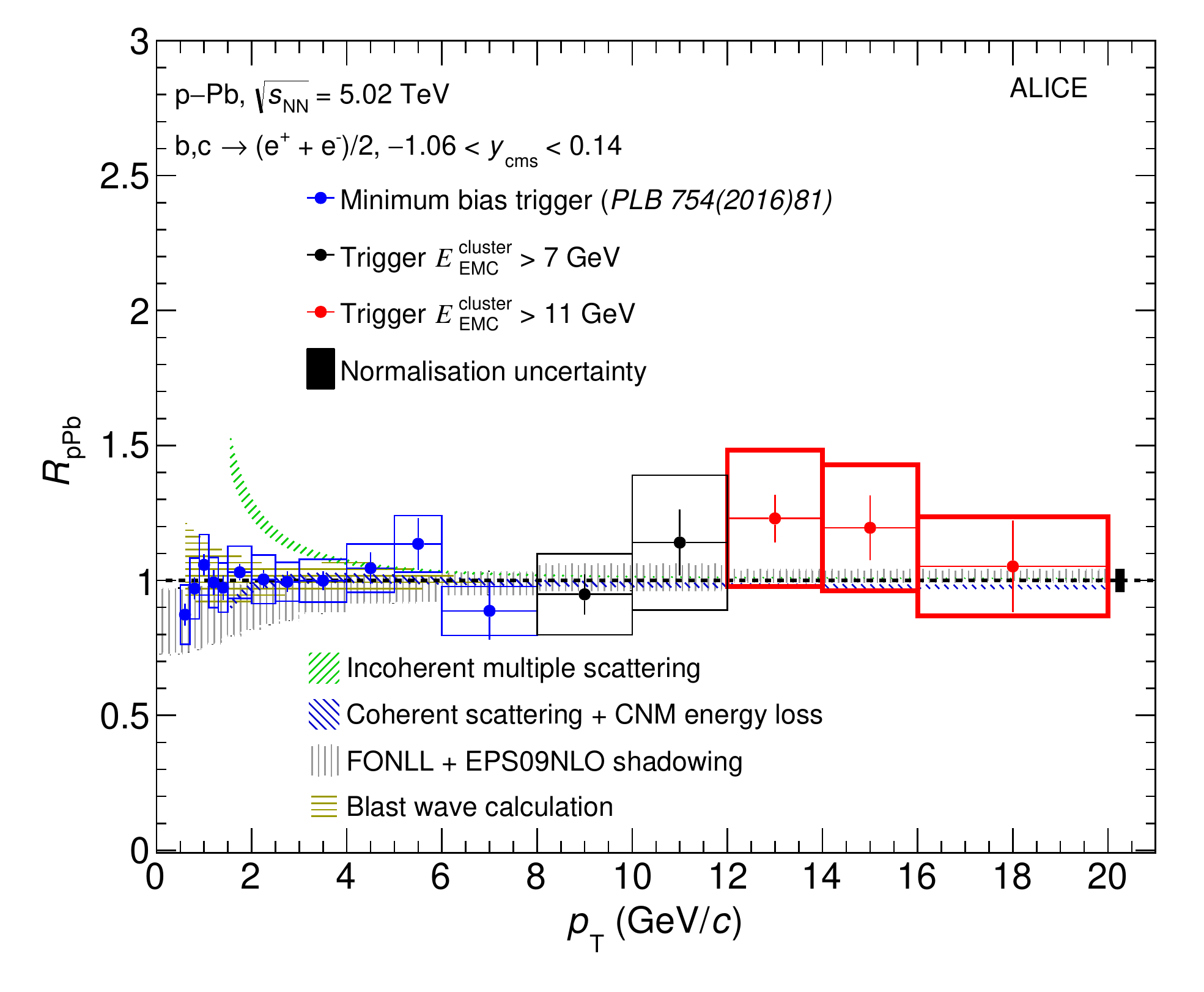}
\caption{Nuclear modification factor, \RpPb, of electrons from heavy-flavour hadron decays as a function of transverse momentum for minimum-bias p--Pb collisions at $\sqrt{s_{\rm NN}}$ = 5.02 TeV. The vertical bars represent the statistical uncertainties, and the boxes indicate the systematic uncertainties. The systematic uncertainty from the normalisation, common to all points, is shown as a solid box at high \pt at \RpPb = 1. The published results~\cite{Adam:2015qda} are updated using the heavy-flavour hadron decays measurement obtained by ALICE in pp collisions at $\sqrt{s}$ = 5.02 TeV~\cite{Acharya:2019mom}. The points above 8 \GeVc are updated and extended using the EMCal trigger. The results are compared with theoretical models~\cite{Kang:2014hha, Sharma:2009hn, Sickles:2013yna, Cacciari:1998it, Eskola:2009uj}, as described in the text.}
\label{fig:RpPb}
\end{figure}

Figure~\ref{fig:RpPb} shows the nuclear modification factor \RpPb of electrons from heavy-flavour hadron decays as a function of transverse momentum. The published results for 0.5 $<$ \pt $<$ 8 \GeVc~\cite{Adam:2015qda} are updated using the heavy-flavour hadron decays  measurements obtained by ALICE in pp collisions at $\sqrt{s}$ = 5.02 TeV~\cite{Acharya:2019mom}.
 The results from 8 $<$ \pt $<$ 20 \GeVc were calculated using the \pt-differential invariant cross section obtained by the EMCal trigger, as presented in Fig.~\ref{fig:cross_section1}.
 
The statistical and systematic uncertainties of the spectra in p--Pb and pp collisions were propagated as independent uncertainties. The normalisation uncertainties are shown as a solid box around the dotted line along \RpPb = 1. The \RpPb is consistent with unity within uncertainties over the whole \pt range of the measurement. Thus, the measurements are consistent with no modification over the measured \pt range.
Heavy-flavour electrons coming from beauty decays are dominant in the high-\pt region, in particular for \pt $>$ 4 GeV/c~\cite{Adam:2014qja, Abelev:2012sca}, where the measurements were extended with the EMCal trigger. The results thus show that the beauty production is not modified in p--Pb collisions within the kinematic range of this measurement. 

The results are compared with different theoretical models. 
 Theoretical model calculations which consider coherent multiple scatterings, including energy loss in the CNM and nuclear shadowing~~\cite{Sharma:2009hn},  results from pQCD calculations, using FONLL~\cite{Cacciari:1998it} + EPS09NLO~\cite{Eskola:2009uj}, that include initial-state effects (nuclear shadowing), and 
Blast-wave calculations~\cite{Sickles:2013yna}, which assume the formation of a hydrodynamically expanding medium, are all in agreement with the measurements, predicting \RpPb close to unity.
Calculations based on incoherent multiple scatterings predict an enhancement at low \pt~\cite{Kang:2014hha}, which is not observed in the measurements.

\begin{figure}[h!]
\centering
\includegraphics[width=0.9\linewidth]{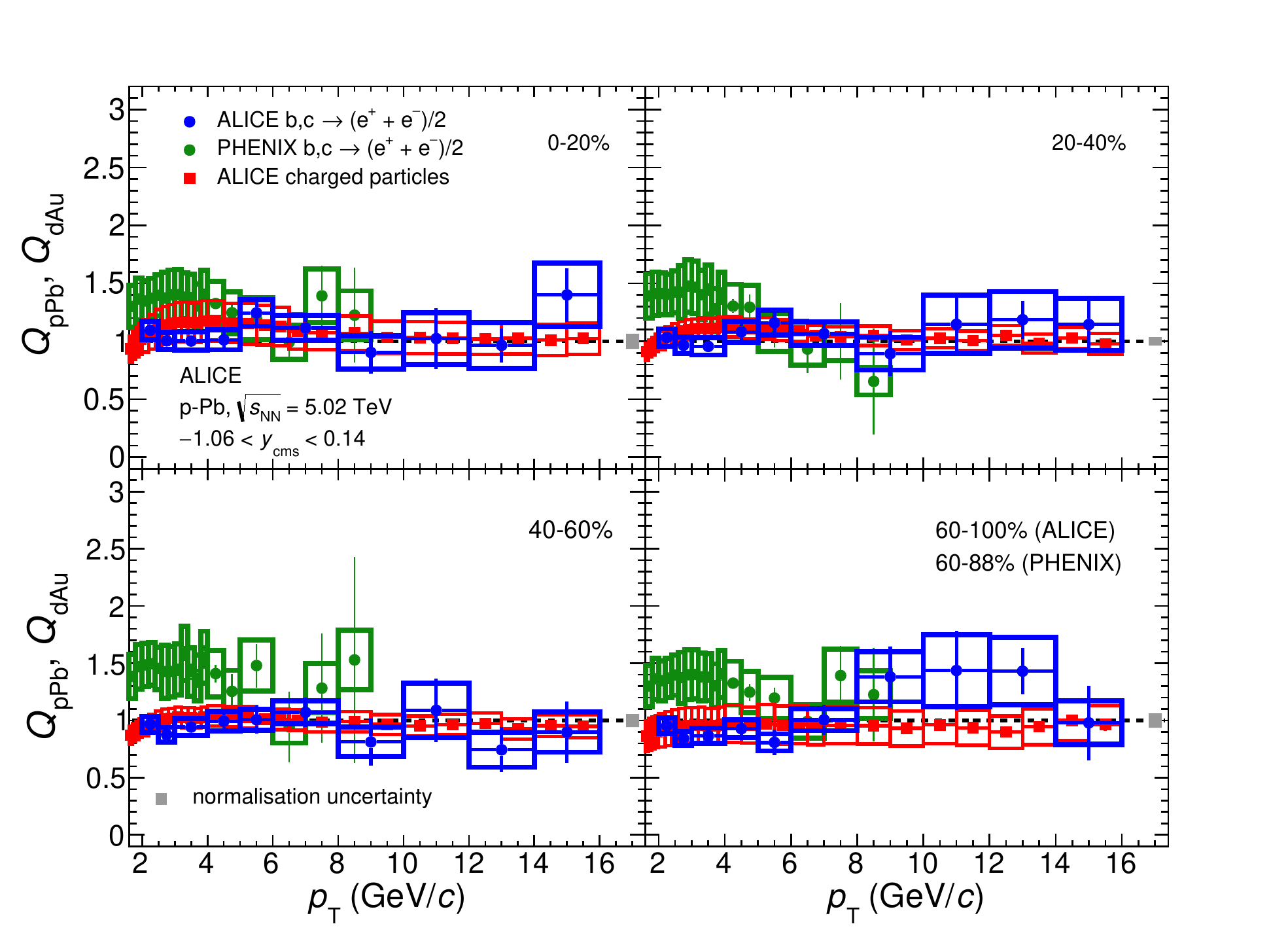}
\caption{Nuclear modification factors $Q_{\rm pPb}$ as a function of \pt in the 0-20\%, 20-40\%, 40-60\%, and 60-100\% multiplicity classes selected with the ZNA estimator in \pPb collisions at $\sqrt{s_{\rm NN}}$ = 5.02 TeV. The different panels of the figure are for different multiplicity classes. The vertical error bars and the empty boxes represent the statistical and systematic uncertainties, respectively. The solid boxes at high \pt at $Q_{\rm pPb}$ = 1 represent the normalisation uncertainties. The results are compared with the PHENIX results on electrons from heavy-flavour hadron decays~\cite{Adare:2012yxa} in d$+$Au collisions at $\sqrt{s_{\rm NN}}$ = 200 GeV and with ALICE charged particle results~\cite{Adam:2014qja} in \pPb collisions at $\sqrt{s_{\rm NN}}$ = 5.02~TeV.}
\label{fig:QpPb}
\end{figure}

The multiplicity dependence of the production of heavy-flavour electrons was studied by measuring the nuclear modification factor in each multiplicity class, $Q_{\rm pPb}$, which was calculated as defined in Eq.~\ref{eq:QpPb}.
Figure~\ref{fig:QpPb} shows the $Q_{\rm pPb}$ results for 0-20\%, 20-40\%, 40-60\%, and 60-100\% multiplicity classes in p--Pb collisions at $\sqrt{s_{\rm NN}}$ = 5.02 TeV. The uncertainty on the average nuclear overlap function $\langle T_{\rm pPb}^{\rm mult} \rangle$ for each centrality selection is given in Tab.~\ref{TableTpPb}. The pp reference uncertainties were propagated to the final uncertainty of $Q_{\rm pPb}$. It is found that the $Q_{\rm pPb}$ is close to unity. A comparison between these results and the PHENIX measurements of electrons from heavy-quark decays in d+Au collisions at $\sqrt{s_{\rm NN}}$ = 200~GeV~\cite{Adare:2012yxa} is shown in Fig.~\ref{fig:QpPb}. This figure also shows the ALICE results for charged particles measured in \pPb collisions at  $\sqrt{s_{\rm NN}}$ = 5.02 TeV~\cite{Adam:2014qja}.

These measurements are compatible with charged-particle results, which may hint to no mass dependence of particle production in \pPb collisions. However, PHENIX results are higher than these results, which may indicate smaller CNM effects at the LHC.
The differences can also be explained by the fact that the radial flow at RHIC is expected to be larger than the radial flow at the LHC~\cite{Sickles:2013yna, Adare:2012yxa}.

 In Pb--Pb collisions, a suppression of electrons from heavy-flavour hadron decays was observed not only in the 0-10\% most central but also in the 50-80\% centrality class~\cite{Adam:2016khe}, and the magnitude of the suppression increases from peripheral to the most central collisions. On the other hand, in p--Pb collisions, the $Q_{\rm pPb}$ is consistent with unity within the statistical and systematic uncertainties over the whole \pt range of the measurement, showing no evidence for a multiplicity dependence. The spectrum of electrons from heavy-flavour hadron decays in p--Pb collisions is thus consistent with the spectrum in pp collisions at the same centre-of-mass energy scaled by the number of binary collisions for all centrality bins.
 
The ratio of the nuclear modification factor of electrons from heavy-flavour hadron decays in central collisions with respect to peripheral collisions was calculated as defined in Eq.~\ref{eq:Qcp}.

\begin{figure}[h!]
\centering
\includegraphics[width=0.6\linewidth]{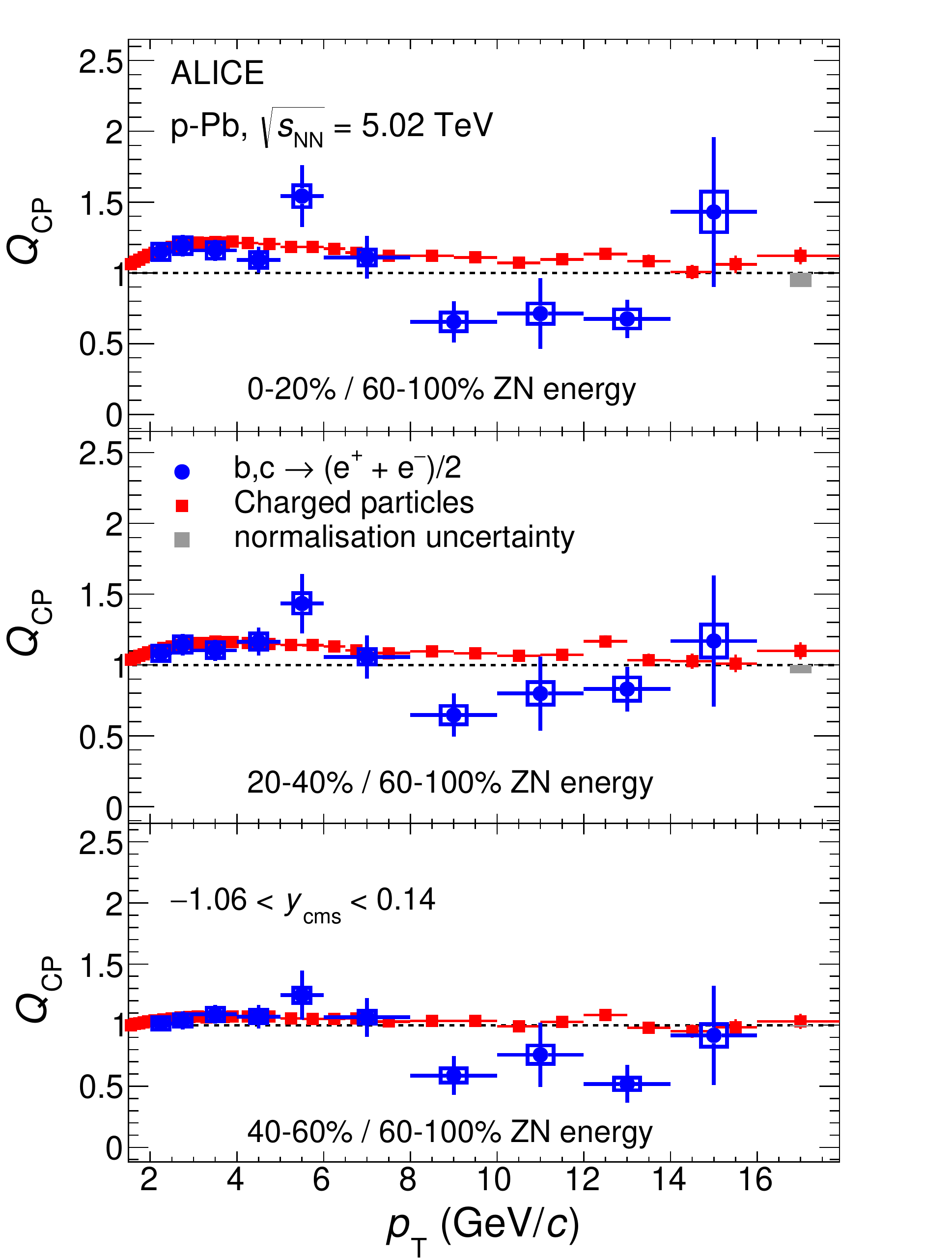}
\caption{$Q_{\rm cp}$ of electrons from heavy-flavour hadron decays in 0-20\%, 20-40\% and 40-60\% multiplicity classes in p--Pb collisions at $\sqrt{s_{\rm NN}}$ = 5.02 TeV. The vertical error bars and the empty boxes represent the statistical and systematic uncertainties, respectively. The solid boxes at high \pt at $Q_{\rm cp}$ = 1 represent the normalisation uncertainties. The results are compared to ALICE results on charged particles in p--Pb collisions at $\sqrt{s_{\rm NN}}$ = 5.02 TeV ~\cite{Adam:2014qja}.}
\label{fig:Qcp}
\end{figure}

The advantage of measuring the $Q_{\rm cp}$ is that it has a smaller systematic uncertainty when compared to $Q_{\rm pPb}$, since $Q_{\rm cp}$ does not depend on the pp reference. Also, some of the uncertainties are correlated for different centralities and they cancel when considering the ratios. 
Figure~\ref{fig:Qcp} shows the $Q_{\rm cp}$ of electrons from heavy-flavour hadron decays in \pPb collisions at $\sqrt{s_{\rm NN}}$ = 5.02 TeV. 
The results are consistent with unity given the statistical and systematic uncertainties. 
A comparison of the electrons from heavy-flavour hadron decays and charged particles $Q_{\rm cp}$ is shown in Fig.~\ref{fig:Qcp}.
Within systematic uncertainties, the results are compatible and no conclusion about mass dependence can be obtained.  

\FloatBarrier

Since the $Q_{\rm cp}$ results are compatible with unity within systematic uncertainties, no modification of the spectra in central collisions with respect to peripheral collisions is observed. This feature is an indication that CNM effects in the production of electrons from heavy-flavour hadron decays within the measured \pt interval are not centrality dependent. 

Several previous measurements of light-particle production show that \pPb collisions cannot be explained by an incoherent superposition of pp collisions, but show the presence of coherent and collective effects~\cite{Abelev:2012ola, Abelev:2013bla, Abelev:2013haa, ABELEV:2013wsa}. For light flavours, there is an indication of Cronin enhancement in the results for central collisions, although the results are also compatible with unity, given the normalisation systematic uncertainty. Our measurements probed such effects in the heavy-flavour sector,  showing that these effects are similar for all centralities within the uncertainties.

\section{Summary}\label{summary}

The \pt-differential cross sections of electrons from heavy-flavour hadron decays were measured up to  20 \GeVc using EMCal triggered data, which extends the previously reported ALICE measurement~\cite{Adam:2015qda}.
It is found that the updated and extended measurement of the \RpPb is consistent with unity as observed in ~\cite{Adam:2015qda} and at the same time is still consistent with 
theoretical predictions including CNM effects and radial flow. 
The \pt-differential cross sections of electrons from heavy-flavour hadron decays were also measured in four multiplicity classes in \pPb collisions at $\sqrt{s_{\rm NN}}$ = 5.02 TeV in the transverse momentum range 2 $< p_{\rm T} <$ 16 \GeVc at midrapidity. 
The TPC detector was used to measure the yield for 2 $< p_{\rm T} <$ 8 \GeVc and the combination of the TPC and the EMCal detectors were used for 8 $< p_{\rm T} <$ 16 \GeVc in EMCal triggered data set.
The nuclear modification factor, $Q_{\rm pPb}$, was evaluated for four multiplicity classes and the results are all consistent with unity. 
There is no indication of multiplicity dependence in the production of electrons from heavy-flavour decays  in \pPb collisions with respect to that of pp collisions at the same centre-of-mass energy. 
The $Q_{\rm cp}$ results are consistent  with unity with smaller statistical and systematic uncertainties, showing that the production of electrons from heavy-flavour hadron decays is the same for central and peripheral collisions.
The $Q_{\rm pPb}$ and $Q_{\rm cp}$ measurements suggest that there is no multiplicity dependence of the production of electrons from heavy-flavour hadron decays in \pPb collisions at $\sqrt{s_{\rm NN}}$ = 5.02 TeV. Hence, their production is not affected by the number of charged particles produced in the collision.  
These results indicate that the suppression of the yield of heavy-flavour production in Pb--Pb collisions at high \pt is not an initial-state effect, but a final-state effect induced by the hot medium.
They also indicate that the CNM effects on heavy-flavour production are negligible in both central and peripheral collisions at midrapidity.

\newenvironment{acknowledgement}{\relax}{\relax}
\begin{acknowledgement}
\section*{Acknowledgements}

The ALICE Collaboration would like to thank all its engineers and technicians for their invaluable contributions to the construction of the experiment and the CERN accelerator teams for the outstanding performance of the LHC complex.
The ALICE Collaboration gratefully acknowledges the resources and support provided by all Grid centres and the Worldwide LHC Computing Grid (WLCG) collaboration.
The ALICE Collaboration acknowledges the following funding agencies for their support in building and running the ALICE detector:
A. I. Alikhanyan National Science Laboratory (Yerevan Physics Institute) Foundation (ANSL), State Committee of Science and World Federation of Scientists (WFS), Armenia;
Austrian Academy of Sciences, Austrian Science Fund (FWF): [M 2467-N36] and Nationalstiftung f\"{u}r Forschung, Technologie und Entwicklung, Austria;
Ministry of Communications and High Technologies, National Nuclear Research Center, Azerbaijan;
Conselho Nacional de Desenvolvimento Cient\'{\i}fico e Tecnol\'{o}gico (CNPq), Financiadora de Estudos e Projetos (Finep), Funda\c{c}\~{a}o de Amparo \`{a} Pesquisa do Estado de S\~{a}o Paulo (FAPESP) and Universidade Federal do Rio Grande do Sul (UFRGS), Brazil;
Ministry of Education of China (MOEC) , Ministry of Science \& Technology of China (MSTC) and National Natural Science Foundation of China (NSFC), China;
Ministry of Science and Education and Croatian Science Foundation, Croatia;
Centro de Aplicaciones Tecnol\'{o}gicas y Desarrollo Nuclear (CEADEN), Cubaenerg\'{\i}a, Cuba;
Ministry of Education, Youth and Sports of the Czech Republic, Czech Republic;
The Danish Council for Independent Research | Natural Sciences, the VILLUM FONDEN and Danish National Research Foundation (DNRF), Denmark;
Helsinki Institute of Physics (HIP), Finland;
Commissariat \`{a} l'Energie Atomique (CEA), Institut National de Physique Nucl\'{e}aire et de Physique des Particules (IN2P3) and Centre National de la Recherche Scientifique (CNRS) and R\'{e}gion des  Pays de la Loire, France;
Bundesministerium f\"{u}r Bildung und Forschung (BMBF) and GSI Helmholtzzentrum f\"{u}r Schwerionenforschung GmbH, Germany;
General Secretariat for Research and Technology, Ministry of Education, Research and Religions, Greece;
National Research, Development and Innovation Office, Hungary;
Department of Atomic Energy Government of India (DAE), Department of Science and Technology, Government of India (DST), University Grants Commission, Government of India (UGC) and Council of Scientific and Industrial Research (CSIR), India;
Indonesian Institute of Science, Indonesia;
Centro Fermi - Museo Storico della Fisica e Centro Studi e Ricerche Enrico Fermi and Istituto Nazionale di Fisica Nucleare (INFN), Italy;
Institute for Innovative Science and Technology , Nagasaki Institute of Applied Science (IIST), Japanese Ministry of Education, Culture, Sports, Science and Technology (MEXT) and Japan Society for the Promotion of Science (JSPS) KAKENHI, Japan;
Consejo Nacional de Ciencia (CONACYT) y Tecnolog\'{i}a, through Fondo de Cooperaci\'{o}n Internacional en Ciencia y Tecnolog\'{i}a (FONCICYT) and Direcci\'{o}n General de Asuntos del Personal Academico (DGAPA), Mexico;
Nederlandse Organisatie voor Wetenschappelijk Onderzoek (NWO), Netherlands;
The Research Council of Norway, Norway;
Commission on Science and Technology for Sustainable Development in the South (COMSATS), Pakistan;
Pontificia Universidad Cat\'{o}lica del Per\'{u}, Peru;
Ministry of Science and Higher Education and National Science Centre, Poland;
Korea Institute of Science and Technology Information and National Research Foundation of Korea (NRF), Republic of Korea;
Ministry of Education and Scientific Research, Institute of Atomic Physics and Ministry of Research and Innovation and Institute of Atomic Physics, Romania;
Joint Institute for Nuclear Research (JINR), Ministry of Education and Science of the Russian Federation, National Research Centre Kurchatov Institute, Russian Science Foundation and Russian Foundation for Basic Research, Russia;
Ministry of Education, Science, Research and Sport of the Slovak Republic, Slovakia;
National Research Foundation of South Africa, South Africa;
Swedish Research Council (VR) and Knut \& Alice Wallenberg Foundation (KAW), Sweden;
European Organization for Nuclear Research, Switzerland;
Suranaree University of Technology (SUT), National Science and Technology Development Agency (NSDTA) and Office of the Higher Education Commission under NRU project of Thailand, Thailand;
Turkish Atomic Energy Agency (TAEK), Turkey;
National Academy of  Sciences of Ukraine, Ukraine;
Science and Technology Facilities Council (STFC), United Kingdom;
National Science Foundation of the United States of America (NSF) and United States Department of Energy, Office of Nuclear Physics (DOE NP), United States of America.    
\end{acknowledgement}

\bibliographystyle{utphys}

\bibliography{hfe_pPb_multiplicity}

\newpage
\appendix
\section{The ALICE Collaboration}
\label{app:collab}

\begingroup
\small
\begin{flushleft}
S.~Acharya\Irefn{org141}\And 
D.~Adamov\'{a}\Irefn{org93}\And 
S.P.~Adhya\Irefn{org141}\And 
A.~Adler\Irefn{org74}\And 
J.~Adolfsson\Irefn{org80}\And 
M.M.~Aggarwal\Irefn{org98}\And 
G.~Aglieri Rinella\Irefn{org34}\And 
M.~Agnello\Irefn{org31}\And 
N.~Agrawal\Irefn{org10}\And 
Z.~Ahammed\Irefn{org141}
% ...
\And
\renewcommand\labelenumi{\textsuperscript{\theenumi}~}

\section*{Affiliation notes}
\renewcommand\theenumi{\roman{enumi}}
\begin{Authlist}
\item \Adef{org*}Deceased
\item \Adef{orgI}Dipartimento DET del Politecnico di Torino, Turin, Italy
\item \Adef{orgII}M.V. Lomonosov Moscow State University, D.V. Skobeltsyn Institute of Nuclear, Physics, Moscow, Russia
\item \Adef{orgIII}Department of Applied Physics, Aligarh Muslim University, Aligarh, India
\item \Adef{orgIV}Institute of Theoretical Physics, University of Wroclaw, Poland
\end{Authlist}

\section*{Collaboration Institutes}
\renewcommand\theenumi{\arabic{enumi}~}
\begin{Authlist}
\item \Idef{org1}A.I. Alikhanyan National Science Laboratory (Yerevan Physics Institute) Foundation, Yerevan, Armenia
\item \Idef{org2}Bogolyubov Institute for Theoretical Physics, National Academy of Sciences of Ukraine, Kiev, Ukraine
\item \Idef{org3}Bose Institute, Department of Physics  and Centre for Astroparticle Physics and Space Science (CAPSS), Kolkata, India
\item \Idef{org4}Budker Institute for Nuclear Physics, Novosibirsk, Russia
\item \Idef{org5}California Polytechnic State University, San Luis Obispo, California, United States
\item \Idef{org6}Central China Normal University, Wuhan, China
\item \Idef{org7}Centre de Calcul de l'IN2P3, Villeurbanne, Lyon, France
\item \Idef{org8}Centro de Aplicaciones Tecnol\'{o}gicas y Desarrollo Nuclear (CEADEN), Havana, Cuba
\item \Idef{org9}Centro de Investigaci\'{o}n y de Estudios Avanzados (CINVESTAV), Mexico City and M\'{e}rida, Mexico
\item \Idef{org10}Centro Fermi - Museo Storico della Fisica e Centro Studi e Ricerche ``Enrico Fermi', Rome, Italy
\item \Idef{org11}Chicago State University, Chicago, Illinois, United States
\item \Idef{org12}China Institute of Atomic Energy, Beijing, China
\item \Idef{org13}Chonbuk National University, Jeonju, Republic of Korea
\item \Idef{org14}Comenius University Bratislava, Faculty of Mathematics, Physics and Informatics, Bratislava, Slovakia
\item \Idef{org15}COMSATS University Islamabad, Islamabad, Pakistan
\item \Idef{org16}Creighton University, Omaha, Nebraska, United States
\item \Idef{org17}Department of Physics, Aligarh Muslim University, Aligarh, India
\item \Idef{org18}Department of Physics, Pusan National University, Pusan, Republic of Korea
\item \Idef{org19}Department of Physics, Sejong University, Seoul, Republic of Korea
\item \Idef{org20}Department of Physics, University of California, Berkeley, California, United States
\item \Idef{org21}Department of Physics, University of Oslo, Oslo, Norway
\item \Idef{org22}Department of Physics and Technology, University of Bergen, Bergen, Norway
\item \Idef{org23}Dipartimento di Fisica dell'Universit\`{a} 'La Sapienza' and Sezione INFN, Rome, Italy
\item \Idef{org24}Dipartimento di Fisica dell'Universit\`{a} and Sezione INFN, Cagliari, Italy
\item \Idef{org25}Dipartimento di Fisica dell'Universit\`{a} and Sezione INFN, Trieste, Italy
\item \Idef{org26}Dipartimento di Fisica dell'Universit\`{a} and Sezione INFN, Turin, Italy
\item \Idef{org27}Dipartimento di Fisica e Astronomia dell'Universit\`{a} and Sezione INFN, Bologna, Italy
\item \Idef{org28}Dipartimento di Fisica e Astronomia dell'Universit\`{a} and Sezione INFN, Catania, Italy
\item \Idef{org29}Dipartimento di Fisica e Astronomia dell'Universit\`{a} and Sezione INFN, Padova, Italy
\item \Idef{org30}Dipartimento di Fisica `E.R.~Caianiello' dell'Universit\`{a} and Gruppo Collegato INFN, Salerno, Italy
\item \Idef{org31}Dipartimento DISAT del Politecnico and Sezione INFN, Turin, Italy
\item \Idef{org32}Dipartimento di Scienze e Innovazione Tecnologica dell'Universit\`{a} del Piemonte Orientale and INFN Sezione di Torino, Alessandria, Italy
\item \Idef{org33}Dipartimento Interateneo di Fisica `M.~Merlin' and Sezione INFN, Bari, Italy
\item \Idef{org34}European Organization for Nuclear Research (CERN), Geneva, Switzerland
\item \Idef{org35}Faculty of Electrical Engineering, Mechanical Engineering and Naval Architecture, University of Split, Split, Croatia
\item \Idef{org36}Faculty of Engineering and Science, Western Norway University of Applied Sciences, Bergen, Norway
\item \Idef{org37}Faculty of Nuclear Sciences and Physical Engineering, Czech Technical University in Prague, Prague, Czech Republic
\item \Idef{org38}Faculty of Science, P.J.~\v{S}af\'{a}rik University, Ko\v{s}ice, Slovakia
\item \Idef{org39}Frankfurt Institute for Advanced Studies, Johann Wolfgang Goethe-Universit\"{a}t Frankfurt, Frankfurt, Germany
\item \Idef{org40}Gangneung-Wonju National University, Gangneung, Republic of Korea
\item \Idef{org41}Gauhati University, Department of Physics, Guwahati, India
\item \Idef{org42}Helmholtz-Institut f\"{u}r Strahlen- und Kernphysik, Rheinische Friedrich-Wilhelms-Universit\"{a}t Bonn, Bonn, Germany
\item \Idef{org43}Helsinki Institute of Physics (HIP), Helsinki, Finland
\item \Idef{org44}High Energy Physics Group,  Universidad Aut\'{o}noma de Puebla, Puebla, Mexico
\item \Idef{org45}Hiroshima University, Hiroshima, Japan
\item \Idef{org46}Hochschule Worms, Zentrum  f\"{u}r Technologietransfer und Telekommunikation (ZTT), Worms, Germany
\item \Idef{org47}Horia Hulubei National Institute of Physics and Nuclear Engineering, Bucharest, Romania
\item \Idef{org48}Indian Institute of Technology Bombay (IIT), Mumbai, India
\item \Idef{org49}Indian Institute of Technology Indore, Indore, India
\item \Idef{org50}Indonesian Institute of Sciences, Jakarta, Indonesia
\item \Idef{org51}INFN, Laboratori Nazionali di Frascati, Frascati, Italy
\item \Idef{org52}INFN, Sezione di Bari, Bari, Italy
\item \Idef{org53}INFN, Sezione di Bologna, Bologna, Italy
\item \Idef{org54}INFN, Sezione di Cagliari, Cagliari, Italy
\item \Idef{org55}INFN, Sezione di Catania, Catania, Italy
\item \Idef{org56}INFN, Sezione di Padova, Padova, Italy
\item \Idef{org57}INFN, Sezione di Roma, Rome, Italy
\item \Idef{org58}INFN, Sezione di Torino, Turin, Italy
\item \Idef{org59}INFN, Sezione di Trieste, Trieste, Italy
\item \Idef{org60}Inha University, Incheon, Republic of Korea
\item \Idef{org61}Institut de Physique Nucl\'{e}aire d'Orsay (IPNO), Institut National de Physique Nucl\'{e}aire et de Physique des Particules (IN2P3/CNRS), Universit\'{e} de Paris-Sud, Universit\'{e} Paris-Saclay, Orsay, France
\item \Idef{org62}Institute for Nuclear Research, Academy of Sciences, Moscow, Russia
\item \Idef{org63}Institute for Subatomic Physics, Utrecht University/Nikhef, Utrecht, Netherlands
\item \Idef{org64}Institute for Theoretical and Experimental Physics, Moscow, Russia
\item \Idef{org65}Institute of Experimental Physics, Slovak Academy of Sciences, Ko\v{s}ice, Slovakia
\item \Idef{org66}Institute of Physics, Homi Bhabha National Institute, Bhubaneswar, India
\item \Idef{org67}Institute of Physics of the Czech Academy of Sciences, Prague, Czech Republic
\item \Idef{org68}Institute of Space Science (ISS), Bucharest, Romania
\item \Idef{org69}Institut f\"{u}r Kernphysik, Johann Wolfgang Goethe-Universit\"{a}t Frankfurt, Frankfurt, Germany
\item \Idef{org70}Instituto de Ciencias Nucleares, Universidad Nacional Aut\'{o}noma de M\'{e}xico, Mexico City, Mexico
\item \Idef{org71}Instituto de F\'{i}sica, Universidade Federal do Rio Grande do Sul (UFRGS), Porto Alegre, Brazil
\item \Idef{org72}Instituto de F\'{\i}sica, Universidad Nacional Aut\'{o}noma de M\'{e}xico, Mexico City, Mexico
\item \Idef{org73}iThemba LABS, National Research Foundation, Somerset West, South Africa
\item \Idef{org74}Johann-Wolfgang-Goethe Universit\"{a}t Frankfurt Institut f\"{u}r Informatik, Fachbereich Informatik und Mathematik, Frankfurt, Germany
\item \Idef{org75}Joint Institute for Nuclear Research (JINR), Dubna, Russia
\item \Idef{org76}Korea Institute of Science and Technology Information, Daejeon, Republic of Korea
\item \Idef{org77}KTO Karatay University, Konya, Turkey
\item \Idef{org78}Laboratoire de Physique Subatomique et de Cosmologie, Universit\'{e} Grenoble-Alpes, CNRS-IN2P3, Grenoble, France
\item \Idef{org79}Lawrence Berkeley National Laboratory, Berkeley, California, United States
\item \Idef{org80}Lund University Department of Physics, Division of Particle Physics, Lund, Sweden
\item \Idef{org81}Nagasaki Institute of Applied Science, Nagasaki, Japan
\item \Idef{org82}Nara Women{'}s University (NWU), Nara, Japan
\item \Idef{org83}National and Kapodistrian University of Athens, School of Science, Department of Physics , Athens, Greece
\item \Idef{org84}National Centre for Nuclear Research, Warsaw, Poland
\item \Idef{org85}National Institute of Science Education and Research, Homi Bhabha National Institute, Jatni, India
\item \Idef{org86}National Nuclear Research Center, Baku, Azerbaijan
\item \Idef{org87}National Research Centre Kurchatov Institute, Moscow, Russia
\item \Idef{org88}Niels Bohr Institute, University of Copenhagen, Copenhagen, Denmark
\item \Idef{org89}Nikhef, National institute for subatomic physics, Amsterdam, Netherlands
\item \Idef{org90}NRC Kurchatov Institute IHEP, Protvino, Russia
\item \Idef{org91}NRNU Moscow Engineering Physics Institute, Moscow, Russia
\item \Idef{org92}Nuclear Physics Group, STFC Daresbury Laboratory, Daresbury, United Kingdom
\item \Idef{org93}Nuclear Physics Institute of the Czech Academy of Sciences, \v{R}e\v{z} u Prahy, Czech Republic
\item \Idef{org94}Oak Ridge National Laboratory, Oak Ridge, Tennessee, United States
\item \Idef{org95}Ohio State University, Columbus, Ohio, United States
\item \Idef{org96}Petersburg Nuclear Physics Institute, Gatchina, Russia
\item \Idef{org97}Physics department, Faculty of science, University of Zagreb, Zagreb, Croatia
\item \Idef{org98}Physics Department, Panjab University, Chandigarh, India
\item \Idef{org99}Physics Department, University of Jammu, Jammu, India
\item \Idef{org100}Physics Department, University of Rajasthan, Jaipur, India
\item \Idef{org101}Physikalisches Institut, Eberhard-Karls-Universit\"{a}t T\"{u}bingen, T\"{u}bingen, Germany
\item \Idef{org102}Physikalisches Institut, Ruprecht-Karls-Universit\"{a}t Heidelberg, Heidelberg, Germany
\item \Idef{org103}Physik Department, Technische Universit\"{a}t M\"{u}nchen, Munich, Germany
\item \Idef{org104}Politecnico di Bari, Bari, Italy
\item \Idef{org105}Research Division and ExtreMe Matter Institute EMMI, GSI Helmholtzzentrum f\"ur Schwerionenforschung GmbH, Darmstadt, Germany
\item \Idef{org106}Rudjer Bo\v{s}kovi\'{c} Institute, Zagreb, Croatia
\item \Idef{org107}Russian Federal Nuclear Center (VNIIEF), Sarov, Russia
\item \Idef{org108}Saha Institute of Nuclear Physics, Homi Bhabha National Institute, Kolkata, India
\item \Idef{org109}School of Physics and Astronomy, University of Birmingham, Birmingham, United Kingdom
\item \Idef{org110}Secci\'{o}n F\'{\i}sica, Departamento de Ciencias, Pontificia Universidad Cat\'{o}lica del Per\'{u}, Lima, Peru
\item \Idef{org111}Shanghai Institute of Applied Physics, Shanghai, China
\item \Idef{org112}St. Petersburg State University, St. Petersburg, Russia
\item \Idef{org113}Stefan Meyer Institut f\"{u}r Subatomare Physik (SMI), Vienna, Austria
\item \Idef{org114}SUBATECH, IMT Atlantique, Universit\'{e} de Nantes, CNRS-IN2P3, Nantes, France
\item \Idef{org115}Suranaree University of Technology, Nakhon Ratchasima, Thailand
\item \Idef{org116}Technical University of Ko\v{s}ice, Ko\v{s}ice, Slovakia
\item \Idef{org117}Technische Universit\"{a}t M\"{u}nchen, Excellence Cluster 'Universe', Munich, Germany
\item \Idef{org118}The Henryk Niewodniczanski Institute of Nuclear Physics, Polish Academy of Sciences, Cracow, Poland
\item \Idef{org119}The University of Texas at Austin, Austin, Texas, United States
\item \Idef{org120}Universidad Aut\'{o}noma de Sinaloa, Culiac\'{a}n, Mexico
\item \Idef{org121}Universidade de S\~{a}o Paulo (USP), S\~{a}o Paulo, Brazil
\item \Idef{org122}Universidade Estadual de Campinas (UNICAMP), Campinas, Brazil
\item \Idef{org123}Universidade Federal do ABC, Santo Andre, Brazil
\item \Idef{org124}University College of Southeast Norway, Tonsberg, Norway
\item \Idef{org125}University of Cape Town, Cape Town, South Africa
\item \Idef{org126}University of Houston, Houston, Texas, United States
\item \Idef{org127}University of Jyv\"{a}skyl\"{a}, Jyv\"{a}skyl\"{a}, Finland
\item \Idef{org128}University of Liverpool, Liverpool, United Kingdom
\item \Idef{org129}University of Science and Technology of China, Hefei, China
\item \Idef{org130}University of Tennessee, Knoxville, Tennessee, United States
\item \Idef{org131}University of the Witwatersrand, Johannesburg, South Africa
\item \Idef{org132}University of Tokyo, Tokyo, Japan
\item \Idef{org133}University of Tsukuba, Tsukuba, Japan
\item \Idef{org134}Universit\'{e} Clermont Auvergne, CNRS/IN2P3, LPC, Clermont-Ferrand, France
\item \Idef{org135}Universit\'{e} de Lyon, Universit\'{e} Lyon 1, CNRS/IN2P3, IPN-Lyon, Villeurbanne, Lyon, France
\item \Idef{org136}Universit\'{e} de Strasbourg, CNRS, IPHC UMR 7178, F-67000 Strasbourg, France, Strasbourg, France
\item \Idef{org137}Universit\'{e} Paris-Saclay Centre d'Etudes de Saclay (CEA), IRFU, D\'{e}partment de Physique Nucl\'{e}aire (DPhN), Saclay, France
\item \Idef{org138}Universit\`{a} degli Studi di Foggia, Foggia, Italy
\item \Idef{org139}Universit\`{a} degli Studi di Pavia, Pavia, Italy
\item \Idef{org140}Universit\`{a} di Brescia, Brescia, Italy
\item \Idef{org141}Variable Energy Cyclotron Centre, Homi Bhabha National Institute, Kolkata, India
\item \Idef{org142}Warsaw University of Technology, Warsaw, Poland
\item \Idef{org143}Wayne State University, Detroit, Michigan, United States
\item \Idef{org144}Westf\"{a}lische Wilhelms-Universit\"{a}t M\"{u}nster, Institut f\"{u}r Kernphysik, M\"{u}nster, Germany
\item \Idef{org145}Wigner Research Centre for Physics, Hungarian Academy of Sciences, Budapest, Hungary
\item \Idef{org146}Yale University, New Haven, Connecticut, United States
\item \Idef{org147}Yonsei University, Seoul, Republic of Korea
\end{Authlist}
\endgroup
\end{document}